\documentclass{ws-ijmpa-mod}
\pdfoutput=1
\usepackage[super,compress]{cite}
\usepackage{hyperref}

\newcommand{\be}{\begin{equation}}
\newcommand{\ee}{\end{equation}}

\newcommand{\GeV}{\,{\rm GeV }}
\newcommand{\kpc}{\,{\rm kpc }}
\newcommand{\eV}{\,{\rm eV}}
\newcommand{\TeV}{\,{\rm TeV}}
\newcommand{\cm}{\,{\rm cm}}

\newcommand{\s}{\,{\rm s}}

\begin{document}

\markboth{Alejandro Ibarra, David Tran, Christoph Weniger}
{Decaying Dark Matter}

%
\catchline{}{}{}{}{}
%

\title{INDIRECT SEARCHES FOR DECAYING DARK MATTER}

\author{ALEJANDRO IBARRA}

\address{Physik-Department, Technische Universit\"at M\"unchen, James-Franck-Stra\ss{}e 1\\
85748 Garching, Germany\\ibarra@tum.de}

\author{DAVID TRAN}

\address{School of Physics and Astronomy, University of Minnesota, Tate Lab, 116 Church Street SE\\
Minneapolis, MN 55455, U.S.A.\\
tran@physics.umn.edu}

\author{CHRISTOPH WENIGER}

\address{
GRAPPA Institute, 
University of Amsterdam, Science Park 904\\
1098XH Amsterdam, Netherlands\\
c.weniger@uva.nl }

\maketitle
\begin{history}
\end{history}

\begin{abstract}
  Numerous observations point towards the existence of an unknown elementary
  particle with no electromagnetic interactions, a large population of which
  was presumably produced in the early stages of the history of the Universe.
  This so-called dark matter has survived until the present day, accounting
  for the 26\% of the present energy budget of the Universe. It remains an
  open question whether the particles comprising the dark matter are
  absolutely stable or whether they have a finite but very long lifetime,
  which is a possibility since there is no known general principle
  guaranteeing perfect stability. In this article we review the observational
  limits on the lifetime of dark matter particles with mass in the GeV $-$ TeV
  range using observations of the cosmic fluxes of antimatter, gamma-rays and
  neutrinos. We also examine some theoretically motivated scenarios that
  provide decaying dark matter candidates.
\end{abstract}

\ccode{PACS numbers: 95.35.+d, 98.70.Sa}

\section{Introduction}
\label{sec:Introduction}
Despite many independent pieces of evidence for the existence of dark matter
particles in the Universe~\cite{Bertone:2004pz, Bergstrom:2012fi,
Jungman:1995df}, very little is known about their properties from the point of
view of particle physics: The spin and the parity of the dark matter 
are completely unknown, while the mass, the interaction cross-section with
nuclei and the dark matter lifetime are only very weakly constrained.  Indeed,
the cosmological and astrophysical evidence for dark matter does not require
the dark matter to be absolutely stable but only to be very long-lived, 
with a lifetime much longer than the age of the Universe of about
13.8~Gyr. The search for the decay products of dark matter can be used to
impose upper bounds on the decay width of dark matter into different final
states. Provided that these final states dominate the decay, this implies lower
bounds on the total dark matter lifetime.

None of the massive particles in the Standard Model are guaranteed to be
absolutely stable on kinematical grounds alone since they could all decay into
lighter matter particles and eventually into photons. Nevertheless, we
clearly observe the existence of long-lived particles, the longevity of which we
attribute to the conservation of certain quantum numbers.  \medskip

\begin{table}[pt]
  \tbl{Longest lived particles in the Standard Model of Particle Physics}
  {\begin{tabular}{@{}cccc@{}} \toprule
      Particle & Lifetime & Decay channel &
      Theoretical interpretation \\
      \colrule
      proton & $\tau>8.2\times 10^{33}$ years & $p\rightarrow e^+
      \pi^0$ & Baryon number conservation\\
      electron & $\tau>4.6\times 10^{26}$ years & $e\rightarrow
      \gamma \nu$ & Electric charge conservation\\
      neutrino & $\tau\gtrsim 10^{12}$ years & $\nu\rightarrow
      \gamma\gamma$ & Lorentz symmetry conservation\\
      neutron & $\tau=880.0\pm 0.9$ s & $n\rightarrow p \bar \nu_e e^-$ & Mild breaking of isospin symmetry \\ 
      dark matter & $\tau\gtrsim 10^{10}$ years & $\vphantom{\overset{?}{\rightarrow}}$ ? & ? \\ \botrule
  \end{tabular} \label{ta1}}
\end{table}

Our first example is the \emph{proton}, which could decay, e.g.,~into a positron
and a neutral pion, $p\rightarrow e^+ \pi^0$. Such a decay has never been
observed, despite the great efforts that have been made, resulting in an
impressive lower limit on the proton lifetime of $\tau> 8.2 \times 10^{33}$
years~\cite{Nishino:2009aa}.  We attribute the non-observation of this
decay to the conservation of baryon number, which is an accidental symmetry of
the renormalizable part of the Standard Model Lagrangian, and which could 
in principle be broken by higher dimensional operators. In fact, some well-motivated
extensions of the Standard Model, such as Grand Unified theories or the most
general Minimal Supersymmetric Standard Model (without $R$-parity
conservation) predict the decay of the proton, thus encouraging further improvements
of proton stability bounds. The second example is the \emph{electron},
which could, e.g.,~decay into a neutrino and a photon, $e\rightarrow
\nu\gamma$. This decay mode has been searched for by the Borexino
collaboration, resulting in the lower limit on the electron lifetime of 
$\tau>4.6\times 10^{26}$ years~\cite{Back:2002xz}. The search for this decay 
serves as a test
of electric charge conservation, which on theoretical grounds can be related
to the invariance of the action under phase transformations of the different
complex fields.  The third example is the \emph{neutrino}, which could decay
into two photons, $\nu\rightarrow \gamma \gamma$. The photons produced in this
decay have been searched for in the cosmic microwave background, resulting in
the lower limit $\tau\gtrsim 10^{12}$ years~\cite{Mirizzi:2007jd}. This decay
mode is a test of the conservation of angular momentum, which in turn is
related to the invariance of the action under Lorentz transformations. Our
last example is the \emph{neutron}, which could decay into a proton, an
electron and an electron antineutrino, $n\rightarrow p\bar\nu_e e^-$ (while
preserving baryon number, electric charge and Lorentz symmetry). In contrast
to the first three examples, neutron decay has been observed, with a lifetime
of $\tau=880.0\pm 0.9$ s.\cite{Beringer:1900zz}.  This lifetime is
extraordinarily short compared to the limits on the proton lifetime although
extraordinarily long compared to other strongly interacting particles, a fact
which is  attributed to the mild breaking of the isospin symmetry.  \medskip

While we have a good understanding of why the above four particles are very 
long-lived, and in the case of the electron and the neutrino possibly absolutely
stable, there is no general principle that ensures the absolute stability of
the dark matter particle (see Tab.~\ref{ta1} for a summary). In most models,
dark matter stability is imposed {\it ad hoc} by imposing extra symmetries.
But many well-motivated particle physics models exist which contain unstable,
although very long-lived, dark matter particles. Following the
same rationale as for the proton, it is conceivable that the dark matter
stability could be due to an accidental symmetry of the renormalizable part of
the Lagrangian which is broken by higher dimensional operators,
which could thus induce the dark matter decay. Concretely, for a spin-1/2
dark matter particle, a 3-body decay into Standard Model fermions
could be induced by a dimension six operator suppressed by a large 
mass scale $M$. Then, for ${\cal O}(1)$  couplings, the lifetime can 
be estimated to be~\cite{Eichler:1989br}
\begin{equation}
  \tau_{\rm DM} \sim 10^{26}\,{\rm s}\,\left(\frac{\rm TeV}{m_{\rm DM}}\right)^5
  \left(\frac{M}{10^{15}\,{\rm GeV}}\right)^4\;.
  \label{eq:target}
\end{equation}
Therefore, the search for the decay products of the dark matter particle potentially 
opens a window to physics at very high energies, as large as the Grand Unification
Scale.  \medskip

Assuming that the dark matter indeed has a finite lifetime $\tau_\text{DM}$, the decay 
of dark matter particles with mass $m_{\rm DM}$ produces primary
particles at the point of decay $\vec{r}$, with a rate per unit kinetic energy
$T$ and unit volume given by 
\begin{align} Q(T,\vec
  r)=\displaystyle{\frac{\rho_\text{DM}(\vec{r})}{m_{\rm DM}}\sum_f
  \Gamma_f\frac{dN^f}{dT}} \;, \label{eqn:source} 
\end{align} 
where the sum is over the partial decay rates $\Gamma_f$, $dN^f/dT$ is the energy 
spectrum of the particles produced in the decay channel $f$, and $\rho_{\text{DM}}(\vec{r})$
is the dark matter density at the position $\vec{r}$. 

At cosmological scales, and for our present purposes, we can consider the Universe
as being filled with a homogeneous and isotropic non-relativistic gas of dark
matter particles with a density given by 
\begin{equation}
  \rho_{\text{DM}}(\vec{r})=\Omega_{\rm DM} \rho_c\;,
\end{equation}
where $\Omega_{\rm DM}=0.26$ and $\rho_c=4.9\times 10^{-6}\GeV \cm^{-3}$ is
the critical density of the Universe (we adopt values determined by the Planck
collaboration~\cite{Ade:2013lta}). The distribution of dark matter particles in
the Milky Way is inferred from numerical $N$-body simulations and is not precisely 
known. Some popular choices for the dark matter
density profile, which illustrate the range of uncertainty in the predictions of
the indirect dark matter signatures, are the Navarro-Frenk-White (NFW)
profile~\cite{Navarro:1995iw,Navarro:1996gj}:
\begin{equation}
  \rho_\text{DM}(r)=\frac{\rho_0}{(r/r_\text{s})
  [1+(r/r_\text{s})]^2}\;,
\end{equation}
with scale radius $r_s = 24 ~\rm{kpc}$,~\cite{Cirelli:2010xx} the Einasto
profile~\cite{Navarro:2003ew,Graham:2005xx,Navarro:2008kc}:
\begin{equation}
  \rho_\text{DM}(r)=\rho_0 \exp\left[-\frac{2}{\alpha}
  \left(\frac{r}{r_s}\right)^\alpha\right]\;,
\end{equation}
with $\alpha=0.17$ and $r_s=28~\rm{kpc}$, and the much shallower isothermal
profile~\cite{Bahcall:1980fb}:
\begin{equation}
\rho_{\rm DM}(r) = \frac{\rho_0}{1+r^2/r_s^2}\;,
\end{equation}
with $r_s=4.4~\rm{kpc}$. In all the cases, the overall normalization factor
$\rho_0$ is chosen to reproduce the local dark matter density
$\rho_\odot= 0.39~\text{GeV}/\text{cm}^3$~\cite{Catena:2009mf, Weber:2009pt,
Salucci:2010qr, Pato:2010yq, Iocco:2011jz} with the distance $r_\odot=8.5$~kpc 
of the Sun to the Galactic center.

The decay of dark matter injects energy in form of (anti-)matter, photons and neutrinos into
the intergalactic and interstellar medium, with potential effects on a large
number of cosmological and astrophysical observables.  In general, these
indirect signals exhibit less directional dependence and less amplification
from regions of high dark matter density or at high redshifts than those
associated with self-annihilation processes,  since the production rate is
linear in the dark matter density (as opposed to quadratic in the case of
dark matter self-annihilation).  This leads to subtle differences in search strategies and
in exclusion limits which, in fact, are often weaker for decaying dark matter.
\medskip

In this review we will focus on indirect searches for dark matter signals in
cosmic-ray antimatter, gamma-rays and neutrinos. These observations typically
provide the strongest constraints on the decay of dark matter particles with 
masses in the GeV--TeV range.  This energy range is interesting for at least two
reasons: (1) many of the theoretically well motivated scenarios for decaying
dark matter are related to the electroweak scale and predict dark matter
particles in this mass range, and (2) phenomenologically there is a strong
overlap with indirect searches for WIMPs (weakly interacting massive
particles), and it is important to understand to what extent a WIMP signal could
be clearly discriminated from decaying dark matter. This review is meant to
provide an overview of the relevant experimental constraints and particle
physics models, and we will summarize in a self-contained way how indirect dark 
matter signatures can be calculated. We do not discuss sterile neutrino dark matter 
with keV-scale masses here, which constitutes a somwhat different scenario of decaying 
dark matter. A review of this scenario can be found in Ref.~\citen{Boyarsky:2009ix,Merle:2013gea}. 
Indirect searches for WIMPs are discussed, e.g., in 
Refs.~\citen{Cirelli:2012tf, Bringmann:2012ez}.  \medskip

This review is organized as follows: We start with a discussion of antimatter
signatures, including positrons, antiprotons and antideuterons, in
Section~\ref{sec:anti}. Galactic and cosmological gamma-ray signals will be
discussed in Section~\ref{sec:gamma}, followed by a brief overview of neutrino
searches in Section~\ref{sec:neutrinos}. In Section~\ref{sec:models}, we
discuss a selection of some interesting theoretical scenarios for decaying
dark matter. We finally conclude in Section~\ref{sec:conclusions}.


\section{Antimatter Searches}
\label{sec:anti}
Antimatter particles in the cosmic radiation are an interesting target for
dark matter searches due to the relative rarity of antimatter and the fact that in
typical theoretical scenarios, matter and antimatter are produced in equal amounts 
by dark matter decay (as opposed to ordinary astrophysical processes, which produce
little to no antimatter). Processes involving the creation of primary cosmic rays 
from dark matter decay may thus alter the observed
abundances of antimatter particles in the cosmic radiation in an appreciable
way, thus allowing for indirect detection of dark matter.

Cosmic rays can be divided into two categories: primary cosmic rays
originating in astrophysical sources, presumably supernova remnants, which
accelerate the cosmic-ray particles to high energies; and secondary cosmic 
rays, which are produced
by spallation processes of primary cosmic rays on the interstellar medium.
Dark matter decay may constitute another primary source of primary cosmic 
rays. If the rate of dark matter-induced cosmic-ray production is high enough, 
these fluxes could be observable in the form of a deviation from the expected 
astrophysical background. Even in the case that a dark matter contribution to the cosmic
radiation cannot be clearly identified, the measured fluxes can be used to impose
constraints on particle physics models of dark matter.

In order to predict the locally observable effects of antimatter production
from dark matter decay, one needs to accurately model the propagation of these particles
from their point of production to our position in the Galaxy. Antimatter
particles, after being created by dark matter decay, propagate in a
complicated manner through the Galaxy before reaching the Earth. The most important
effect in the propagation of charged cosmic rays is diffusion. Charged
particles scatter on inhomogeneities of the tangled interstellar magnetic
fields, inducing a random walk-like motion which can be modeled as a diffusion
process. In addition to this, energy losses, drift, annihilation on gas
particles and reacceleration processes can be relevant, depending on the
particular cosmic-ray species.

Antimatter propagation in the Milky Way is commonly described via a stationary
two-zone diffusion model with cylindrical boundary conditions. In this model,
the number density of antiparticles as a function of momentum, position and
time, $f(p,\vec{r},t)$, satisfies the following transport
equation~\cite{Berezinskii, Strong:2007nh}:

\begin{align}
  \frac{\partial}{\partial t} f(p, \vec{r}, t) = {} & Q(p, \vec{r}, t) +
  \vec{\nabla} \cdot (K \vec{\nabla} f - \vec{V}_c f) +
  \frac{\partial}{\partial p} p^2 D_{pp} \frac{\partial}{\partial p}
  \frac{1}{p^2} f \nonumber \\ & {} - \frac{\partial}{\partial p}
  \left[\frac{dp}{dt} f -\frac{p}{3} (\vec{\nabla} \cdot \vec{V}_c) f\right] -
  \frac{1}{\tau_f} f - \frac{1}{\tau_r} f\;.
\end{align}

The terms of the right-hand side correspond to the injection of primary cosmic
rays, the diffusion of cosmic rays due to scattering on magnetic
inhomogeneities, convection by the Galactic wind of particles emitted by the
disk, diffusive reacceleration in momentum space, continuous energy losses,
adiabatic energy loss/gain, and finally losses from fragmentation or
radioactive decay of cosmic rays.

For the propagation of antimatter particles from dark matter decay, the above
propagation equation can be simplified by regarding only stable primary
particle species and neglecting adiabatic energy losses and reacceleration,
which only play an important role at lower energies. Using these
simplifications, the number density of antimatter particles as a function of
kinetic energy\footnote{In the case of cosmic-ray nuclei, $T$ conventionally
refers to the kinetic energy per nucleon.} $T$ is described by the following
diffusion-loss equation, which is valid for electrons/positrons as well as
antiprotons/antideuterons:
\begin{equation}
  0=\frac{\partial f}{\partial t}=
  Q(T,\vec{r}) + \vec{\nabla} \cdot [K(T,\vec{r})\vec{\nabla} f] +
  \frac{\partial}{\partial T} [b(T,\vec{r}) f]
  -\vec{\nabla} \cdot [\vec{V_c}(\vec{r})  f]
  -2 h \delta(z) \Gamma_{\rm ann} f\;.
  \label{eqn:transport}
\end{equation}
We assume free escape boundary conditions, i.e., we take $f(T,\vec{r},t) = 0$
at the boundary of the magnetic diffusion zone, the shape of which is commonly
approximated by a cylinder with half-height in the range $L \simeq
1-15~\rm{kpc}$ and radius $ R \simeq 20 ~\rm{kpc}$.

The first term on the right hand side of the transport equation,
$Q(T,\vec{r})$, is the source term for antiparticles from dark matter decay,
Eq.~(\ref{eqn:source}), which was discussed in Section \ref{sec:Introduction}.
The second term is a diffusion term, which accounts for the propagation of
cosmic rays through the tangled Galactic magnetic fields. The diffusion
coefficient $K(T,\vec{r})$ is often assumed to be constant throughout the
diffusion zone and is usually parametrized in the following
form~\cite{Strong:2007nh}:
\begin{equation}
  K(T)=K_0 \,\beta\, {\cal R}^\delta\,,
\end{equation}
where $\beta\equiv v/c$ with $v$ being the velocity, and ${\cal R}$ is the
rigidity of the particle, which is defined as the momentum in GeV per unit
charge, ${\cal R}\equiv p({\rm GeV})/Z$. The normalization $K_0$ and the
spectral index $\delta$ of the diffusion coefficient are related to the
properties of the interstellar medium and can be determined from measurements
of primary-to-secondary flux ratios of other cosmic-ray species, mainly from
the Boron to Carbon (B/C) ratio~\cite{Maurin:2001sj}. The third term accounts
for energy losses due to inverse Compton scattering on starlight or the cosmic
microwave background, as well as synchrotron radiation and ionization. The 
fourth term is a convection term which accounts for the drift of charged particles 
away from the disk, which is induced by the Milky Way's Galactic wind. This wind has
axial direction and is frequently assumed to be spatially constant inside the diffusion
region: $\vec{V}_c(\vec{r})=V_c\; {\rm sign}(z)\; \vec{e}_z$. The fifth term
accounts for antimatter annihilation with rate $\Gamma_{\rm ann}$, when it
interacts with ordinary matter in the Galactic disk, which is assumed to be an
``infinitely thin'' disk with half-height $h=100$ pc.  
\bigskip

\begin{table}[t]
  \tbl{Astrophysical parameters compatible with the B/C ratio that yield the
  minimal (MIN), median (MED) and maximal (MAX) antiproton fluxes from dark
  matter annihilations; taken from Ref.~\citen{Maurin:2001sj}.}
  {
    \begin{tabular}{ccccc}
      \toprule
      Model & $\delta$ & $K_0\,({\rm kpc}^2/{\rm Myr})$ & $L\,({\rm kpc})$
      & $V_c\,({\rm km}/{\rm s})$ \\
      \colrule
      MIN & 0.85 & 0.0016 & 1 & 13.5 \\
      MED & 0.70 & 0.0112 & 4 & 12 \\
      MAX & 0.46 & 0.0765 & 15 & 5 \\
      \botrule
    \end{tabular}
  }
  \label{tab:param-propagation}
\end{table}

The transport equation, using the parametrizations of the different terms 
given above, has a number of free parameters which have to be determined from
observation. These parameters can be inferred from measurements of flux ratios
of primary and secondary cosmic-ray species, with the Boron-to-Carbon ratio
being the most important.  Because of degeneracies in the impact of the
different parameters on the resulting cosmic-ray fluxes, such observations
cannot determine all parameters independently, resulting in uncertainties in
the prediction of local fluxes, especially when those fluxes originate from
outside the Galactic disk -- as in the case for antimatter from decaying dark
matter, which is created throughout the dark matter halo.  The ranges of the 
astrophysical parameters that are consistent with
the B/C ratio and that produce the minimal (MIN), median (MED) and maximal
(MAX) antimatter fluxes were calculated in Ref.~\citen{Maurin:2001sj} and are
listed in Table \ref{tab:param-propagation}. Note, however, that recent
multiwavelength studies of the latitude profile of synchrotron emission from
cosmic-ray electrons disfavor diffusion zones as thin as
$L\sim1\kpc$~\cite{Bringmann:2011py, DiBernardo:2012zu}.\bigskip

There are different approaches to solving the diffusion-loss equation,
Eq.~(\ref{eqn:transport}). In full generality the transport equation can only
be solved numerically as, for example, in the well-known
\texttt{GALPROP}~\cite{GALPROP} and \texttt{DRAGON}~\cite{DRAGON} codes, which
employ a Crank-Nicolson implicit second-order finite-difference scheme.
Alternatively, the transport equation can be solved semi-analytically by
making certain simplifications and expanding the solution in a series of
trigonometric and Bessel functions and approximating the full solution by a
finite number of terms in the expansion. We discuss these semi-analytical 
solutions in detail in~\ref{apx:propagation}.  \bigskip

Formally, we can write the solution of the transport equation for a particle
species $i$ at the position of the Solar System, $r = r_\odot$, $z = 0$ as as
convolution involving the injection spectrum of antimatter particles and a 
Green's function which describes the effects of cosmic-ray transport,
\begin{equation}
  f_i(T) = \frac{1}{m_\text{DM} \tau_\text{DM}} \int_0^{T_\text{max}} dT' \,
  G_i(T,T') \frac{dN_i(T')}{dT'} \,,
\end{equation}
where $T_\text{max}$ represents the maximum kinetic energy of the antimatter
particles from the decay process.  We also present explicit solutions for the
Green's functions and convenient numerical approximations for the particular
cases of positrons and antiprotons in the appendix.

Given the number density of antimatter particles from dark matter decay as a
result of the transport equation, the flux of primary antiparticles at edge of
the Solar System is given by:
\begin{equation}
  \Phi_i^{\rm{DM}}(T) = \frac{v}{4 \pi} f_i(T).
  \label{flux}
\end{equation}

At energies smaller than $\sim 10$ GeV the antimatter fluxes at the top of the
Earth's atmosphere can differ considerably from the interstellar fluxes due to
solar modulation effects. One frequently used parametrization of the effect of
solar modulation, which can be derived from the full diffusion and convection
equations describing the solar wind, is the force-field
approximation~\cite{1967ApJ149L115G,1968ApJ1541011G}. The fluxes at the top of
the atmosphere in this approximation are related to the interstellar fluxes 
by the following relation~\cite{1987AA184119P}:
\begin{equation}
  \Phi^{\rm TOA}(T_{\rm TOA})=
  \left(
    \frac{2 m T_{\rm TOA}+T_{\rm TOA}^2}{2 m T_{\rm IS}+T_{\rm IS}^2}
  \right)
  \Phi^{\rm IS}(T_{\rm IS}),
  \label{eq:solar-modulation}
\end{equation}
where $m$ is the mass of the cosmic-ray antimatter particle and $T_{\rm IS}=T_{\rm
TOA}+\phi_F$, with $T_{\rm IS}$ and $T_{\rm TOA}$ being the kinetic energies
of the antimatter particles at the heliospheric boundary and at the top of the
Earth's atmosphere, respectively, and $\phi_F$ being the Fisk potential, which
varies between 500 MV and 1.3 GV over the eleven-year solar cycle.

The transport equation simplifies for the particular cases of positrons and
antiprotons since some of the effects can be neglected to good approximation.
We discuss these particular cases in the following.

\subsection{Positrons}
For the case of the positrons, diffusive reacceleration, convection and
annihilations in the Galactic disk can be neglected in the transport equation
in the energy range of interest (above $\sim$ 10 GeV)~\cite{Delahaye:2008ua,
Salati:2007zz}. The transport equation then simplifies to
\begin{equation}
  \vec{\nabla} \cdot [K(E,\vec{r})\vec{\nabla} f_{e^+}] +
  \frac{\partial}{\partial E} [b(E,\vec{r}) f_{e^+}]+Q(E,\vec{r})=0\;,
  \label{transport-positron}
\end{equation}
where we identify the total energy of positrons with the kinetic energy due to
the relative smallness of the electron mass.

The energy loss rate, $b(E,\vec{r})$, is dominated by inverse Compton
scattering (ICS) of the positrons on the interstellar radiation field (ISRF)
and by synchrotron losses on the Galactic $B$-field: $b=b_{\rm ICS}+b_{\rm
syn}$. The part of the energy loss that is due to ICS is given by
\begin{equation}
  b_\text{ICS}(E_e,\vec{r})=\int_0^\infty d\epsilon
  \int_{\epsilon}^{E_\gamma^\text{max}} dE_\gamma\,
  (E_\gamma-\epsilon)\, \frac{d\sigma^{\rm
  IC}(E_e,\epsilon)}{dE_\gamma}f_{\rm ISRF}(\epsilon,\vec{r})\;,
  \label{eq:b-ICS}
\end{equation}
where $f_{\rm ISRF}(\epsilon,\vec{r})$ is the number density of photons of the
interstellar radiation field, which includes the cosmic microwave background,
thermal dust radiation and starlight. An explicit model of the interstellar
radiation field can be found, e.g., in Ref.~\citen{Porter:2005qx}.  For an
electron energy of $E_e = 1\,{\rm GeV}$, $b_\text{ICS}$ ranges between $4.1
\times 10^{-17}\,{\rm GeV}{\rm s}^{-1}$ and $1.9 \times 10^{-15}\,{\rm
GeV}{\rm s}^{-1}$, depending on $\vec{r}$. We see that at higher energies
$b_\text{ICS}$ approximately scales like $\sim E_e^2$. 

The synchrotron energy loss part, on the other hand, is given by
\begin{equation}
  b_\text{syn}(E_e,\vec{r})=\frac{4}{3}\sigma_\text{T}
  \gamma_e^2 \frac{B^2}{2}\;,
  \label{eq:b-syn}
\end{equation}
where $B^2/2$ is the energy density of the Galactic magnetic field, which is
not very well determined.  A conventional choice is $B\simeq
6\,\mu\text{G}\exp(-|z|/2\,{\rm kpc}-r/10\,{\rm kpc})$~\cite{Strong:1998fr}.
At the position of the Sun this magnetic field yields a synchrotron loss rate
of $b_\text{syn}\simeq 4.0\times10^{-17}(E_e/\,{\rm GeV})^2\,{\rm GeV}\;{\rm
s}^{-1}$.

Note that the interaction between the Galactic magnetic field and dark matter
induced electrons and positrons gives rise to synchrotron radiation, which can
be probed by radio observations of the Galactic center and halo. In case of
decaying dark matter, the resulting limits are weaker, however, than those
obtained from local measurements of cosmic rays~\cite{Ishiwata:2008qy,
Zhang:2009pr}.\medskip

A drastic simplification, which greatly simplifies analytical treatments of
the propagation equations, is to assume that the total rate of energy
loss is position-independent and can be parametrized as 
\begin{equation}
  b(E)= \frac{1}{\tau_E} \left(\frac{E}{E_0}\right)^2\;,
\end{equation}
with $E_0=1\,\GeV$ and the time scale $\tau_E\simeq 10^{16}\,\s$, in
accordance with the synchrotron loss rate above. In the relevant energy range,
this approximation is good up to factor of about two to three (see e.g.~Fig.~1
in Ref.~\citen{Zhang:2009ut}).

Rather than measuring the positron flux directly, many experiments measure the
positron fraction, which is less susceptible to systematics since most sources
of systematic error, such as detector acceptance or trigger efficiency, cancel
out when computing the ratio of particle fluxes. The positron fraction is
defined as the flux of positrons divided by the total flux of electrons plus
positrons, and can be calculated as
\begin{eqnarray}
  {\rm PF}(E) &= &\frac{\Phi_{e^+}^{\rm{DM}}(E) + \Phi_{e^+}^{\rm{bkg}}(E)}
  {\Phi^{\rm tot}(E)},
\end{eqnarray}
where the total electron/positron flux is given by
\begin{eqnarray}
  \Phi^{\rm tot}(E)&=&\Phi_{e^-}^{\rm{DM}}(E) +\Phi_{e^+}^{\rm{DM}}(E)
  + \Phi_{e^-}^{\rm{bkg}}(E) + 
  \Phi_{e^+}^{\rm{bkg}}(E)\;,
  \label{eq:tot-electron-flux}
\end{eqnarray}
with $\Phi^{\rm DM}_{e^\pm}$ and $\Phi^{\rm bkg}_{e^\pm}$ being the $e^\pm$
fluxes from dark matter decay and the background fluxes, respectively.  The
background flux of positrons is constituted by secondary positrons produced in
the collision of primary cosmic-ray protons and other nuclei with the
interstellar medium. On the other hand, the background flux of electrons is
constituted by a primary component, presumably produced and accelerated by
supernova remnants, as well as a secondary component, produced by spallation
of cosmic rays on the interstellar medium and which is much smaller than the
primary component. Whereas the spectrum and normalization of secondary
electrons and positrons is calculable in a given propagation model
(e.g.~Ref.~\citen{Moskalenko:1997gh}), the spectrum and normalization of
primary electrons is mainly constrained by the direct measurement.  \medskip

\begin{figure}
  \begin{center}
    \includegraphics[width=0.6\linewidth]{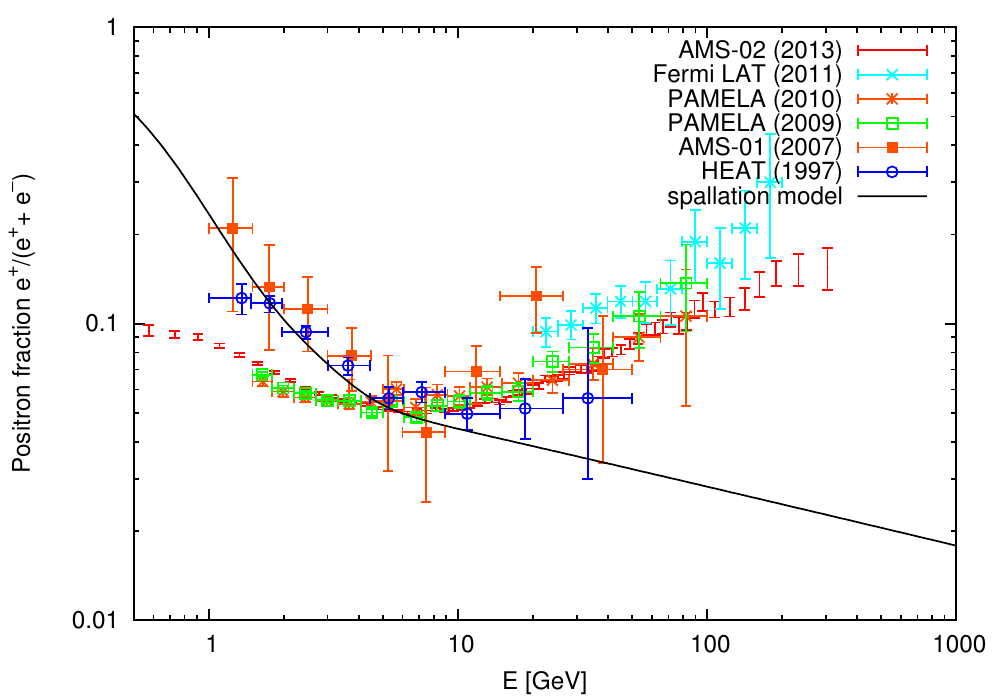}
  \end{center}
  \caption{Recent observations of the positron fraction by
    AMS-02~\cite{Aguilar:2013qda},
    PAMELA~\cite{Adriani:2008zr,Adriani:2010ib}, Fermi
    LAT~\cite{FermiLAT:2011ab}, HEAT~\cite{Barwick:1997ig}. Error bars shown
    include statistical and systematic errors added in quadrature. Also shown
    is a representative theoretical exception from production of positrons by cosmic-ray
    spallation (corresponding to the parametrization in
    Refs.~\citen{Grasso:2009ma, Ibarra:2009dr}). The significant discrepancy
    between theory
    and observation at higher energies is evident. The discrepancy at lower
    energies is due to solar modulation (see Ref.~\citen{Gast:2009xy}).}
  \label{fig:positron_fraction_data}
\end{figure}

The possibility of dark matter contributions to the cosmic-ray positron flux
has attracted a lot of attention in recent years due to the discovery by a
series of increasingly precise cosmic-ray measuments that the positron fraction 
exhibits a steep rise at energies above 10 GeV. Hints of the existence of this 
rise had been observed by AMS-01~\cite{Aguilar:2007yf}, 
HEAT~\cite{Barwick:1997ig,Beatty:2004cy} and
CAPRICE~\cite{Boezio:2000xy}. More recently, the existence of this rise was
confirmed and measured to high precision by PAMELA~\cite{Adriani:2008zr} and
AMS-02~\cite{Aguilar:2013qda}. This behavior is in stark contrast with
conventional models of positron production by cosmic-ray
spallation~\cite{Moskalenko:1997gh}, which predict that the positron fraction
should decrease monotonically with the energy approximately like $\propto
E^{-\delta}$ at energies above a few GeV if the primary spectral indices of
electrons and positrons at injection are similar~\cite{Serpico:2008te}. Recent
observational results obtained by AMS-02~\cite{Aguilar:2013qda},
PAMELA~\cite{Adriani:2008zr,Adriani:2010ib}, Fermi LAT~\cite{FermiLAT:2011ab},
AMS-01~\cite{Aguilar:2007yf} and HEAT~\cite{Barwick:1997ig} are shown in
Fig.~\ref{fig:positron_fraction_data}. Furthermore, recent observations by
Fermi LAT have revealed that the combined flux of electrons and positrons up
to about 1 TeV roughly follows a smooth power law and is harder than expected
from conventional diffusive models~\cite{Abdo:2009zk}. Earlier observations of
a spectral feature in the all-electron spectrum by ATIC~\cite{Chang:2008aa}
and PPB-BETS~\cite{Torii:2008xu} were not confirmed by Fermi LAT. Measurements
by the H.E.S.S. telescope~\cite{Aharonian:2008aa, Aharonian:2009ah}
furthermore indicate that the all-electron spectrum steepens above 1 TeV. Due to
the fact that energetic positrons lose energy efficiently through inverse
Compton scattering, one can infer that the unknown source of positrons must be
local and capable of producing highly energetic cosmic rays.

A number of astrophysical explanations for the positrons excess have been
proposed. Models of $e^+ e^-$ pair production by the interactions of
high-energy photons in the strong magnetic fields of pulsars can reproduce the
positron fraction~\cite{Atoian:1995ux,Hooper:2008kg,Grimani:2004qm}. Other
explanations include a nearby gamma-ray burst~\cite{Ioka:2008cv}, an
inhomogeneous cosmic-ray source distribution~\cite{Shaviv:2009bu}, the
acceleration of secondaries within the sources~\cite{Blasi:2009bd,Ahlers:2009ae} 
or a nearby supernova explosion~\cite{Fujita:2009wk}.

\begin{figure}
  \begin{center}
    \includegraphics[width=0.49\linewidth]{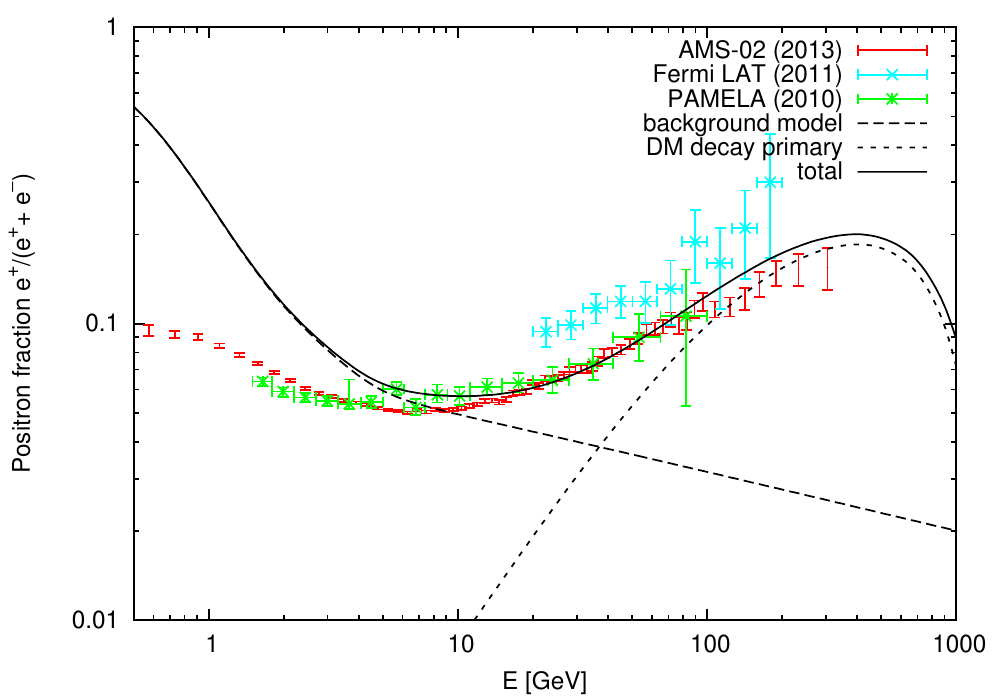}
    \includegraphics[width=0.49\linewidth]{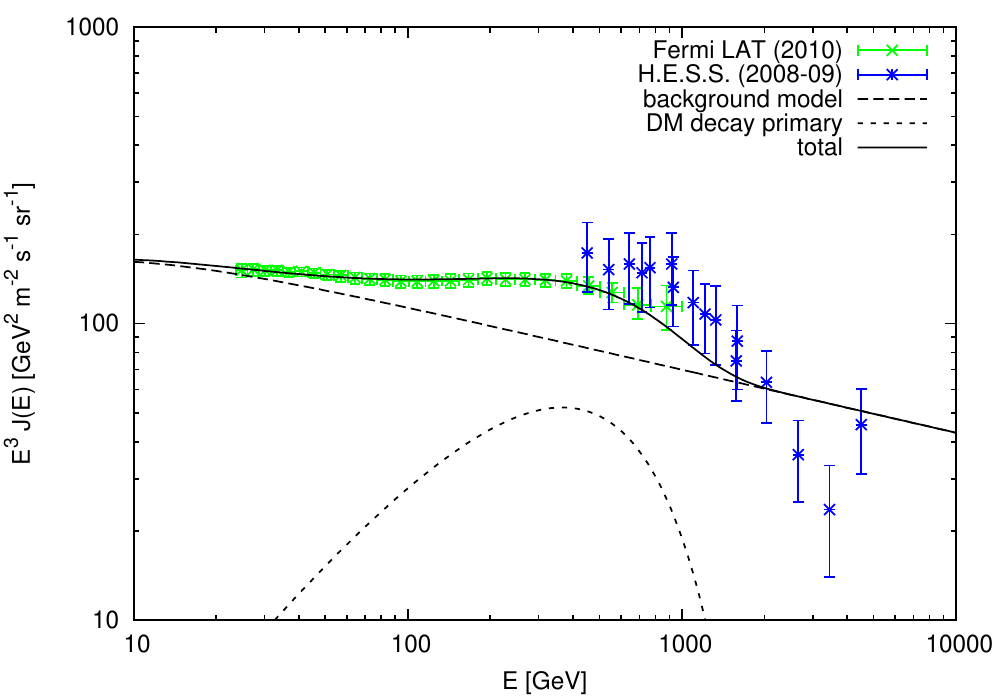}
  \end{center}
  \caption{\emph{Left panel:} Positron fraction for the decay $\psi_\text{DM}
  \to \mu^+ \mu^- \nu$ with $m_\text{DM} = 3000$ GeV and $\tau_\text{DM} =
  1.6 \times 10^{26}$ s.  \emph{Right panel:} All-electron flux for the
  decay $\psi_\text{DM} \to \mu^+ \mu^- \nu$ with $m_\text{DM} = 3000$ GeV
  and $\tau_\text{DM} = 1.6 \times 10^{26}$ s.}
  \label{fig:mumunu_positron_fraction}
\end{figure}

Many authors have explored the more exotic possibility that the positron
excess may be due to dark matter decay into leptonic final states. If the dark matter
particles have sufficiently large mass, their decay could produce highly energetic
positrons and electrons, which might potentially be the origin of the observed anomalies.
Examples of recent works involving decaying dark matter as the source of the
positron excess, or use positron data to derive limits upon the decay rate, include Refs.~\citen{Ibarra:2008jk,Chen:2008yi,
Arvanitaki:2008hq, Ishiwata:2008cu, Ishiwata:2008cv, Chen:2008dh,
Chen:2008qs,  Yin:2008bs, Hamaguchi:2008ta, Pospelov:2008rn,
Arvanitaki:2009yb, Bae:2009bz, Ibarra:2009bm,Cheung:2009si, Chen:2009ew,
Hisano:2009fb, Fukuoka:2009cu, Shirai:2009kh, Choi:2009ng, Kohri:2009yn, Mardon:2009gw,
Arina:2009uq, Bell:2010fk, Carone:2010ha, Ibe:2013jya, Dienes:2013lxa, Jin:2013nta,
Bergstrom:2013jra,Ibarra:2013zia}.  Generally, models of
dark matter which decays mostly into leptons can provide good fits to the
observed lepton abundances provided that the mass of the dark matter is in the
range of a few TeV and the lifetime of the dark matter is around $10^{26} -
10^{27}$~s.~\footnote{Lifetimes of this order exceed the age of the Universe
by a factor of $\sim 10^9$, rendering such dark matter particles nearly
stable on cosmological timescales.} We show a representative example of a fit
of a dark matter signal to the cosmic-ray lepton data in the left panel of
Fig.~\ref{fig:mumunu_positron_fraction}, where we assume that a dark matter
particle of mass $m_\text{DM} = 3$ TeV decays via $\psi_\text{DM} \to \mu^+
\mu^- \nu$ with 100\% branching ratio and lifetime $\tau_\text{DM} = 1.6
\times 10^{26}$ s.  In the right panel of
Fig.~\ref{fig:mumunu_positron_fraction} we display the total
electron-plus-positron flux corresponding to the same set of parameters.  In
Table~\ref{tab:positron_bestfit} we list some of the decay modes which yield
reasonable agreement with the observed positron fraction and total electron
flux, along with the corresponding best-fit values for the dark matter mass
and lifetime. Note that with the advent of AMS-02 results, it becomes
increasingly difficult to obtain satisfactory fits to the
data~\cite{Jin:2013nta} (the fit can be improved, however, by
considering flavor-asymmetric dark matter decays, such as 
${\rm DM}\rightarrow \mu^-\tau^+$~\cite{Feng:2013vva,Masina:2011hu}).

\begin{table}[t]
  \tbl{Sample dark matter decay channels that yield a good numerical fit to the
    positron fraction and the electron flux, together with the best-fit values
    for the dark matter mass and lifetime. Numbers are taken from
    Ref.~\citen{Ibarra:2009dr}.}
  {
    \begin{tabular}{ccc}\toprule
      Decay mode & mass [GeV] & lifetime [$10^{26}$ s] \\\colrule
      $\psi_\text{DM} \to \mu^+ \mu^- \nu$ & 3500 & 1.1 \\
      $\psi_\text{DM} \to \ell^+ \ell^- \nu$ & $2500$ & $1.5$\\
      $\phi_\text{DM} \to \mu^+ \mu^-$ & 2500 & 1.8\\
      \botrule
    \end{tabular}
  }
    \label{tab:positron_bestfit}
\end{table}

Generally, decays into leptons will be accompanied by decays into hadrons,
photons and neutrinos, which allows for complementary tests of dark matter
interpretations of the positron excess in other indirect detection channels.
We discuss these in the following sections.

\subsection{Antiprotons}
Cosmic-ray antiprotons constitute a sensitive probe for exotic -- and usually
charge-symmetric -- contributions from dark matter (with one antiproton
measured per 10,000 protons at 1 GeV kinetic energy~\cite{Adriani:2010rc,Adriani:2012paa}).
Existing models of secondary antiproton production by cosmic-ray spallation match the
observed fluxes well, leaving little room for contributions from exotic 
sources~\cite{Bergstrom:1999jc,Donato:2001ms} such as dark matter. 
Thus, antiproton measurements can be used to impose stringent constraints 
on any dark matter decay modes that involve the production of hadrons.

The general transport equation, Eq.~(\ref{eqn:transport}), can be simplified
for antiprotons by taking into account that in this case energy losses are
negligible due to the relatively large proton mass. The transport equation for
the antiproton density, $f_{\bar p}(T,\vec{r},t)$ then reads
\begin{equation}
  0=\frac{\partial f_{\bar p}}{\partial t}=
  \vec{\nabla} \cdot (K(T,\vec{r})\vec{\nabla} f_{\bar p})
  -\vec{\nabla} \cdot (\vec{V_c}(\vec{r})  f_{\bar p})
  -2 h \delta(z) \Gamma_{\rm ann} f_{\bar p}+Q(T,\vec{r})\;,
  \label{transport-antip}
\end{equation}
where the annihilation rate, $\Gamma_{\rm ann}$, is given by
\begin{equation}
  \Gamma_{\rm ann}=(n_{\rm H}+4^{2/3} n_{\rm He})
  \sigma^{\rm ann}_{\bar p p} v_{\bar p}\;.
\end{equation}
In this expression it has been assumed that the annihilation cross-section
between an antiproton and a helium nucleus is related to the annihilation
cross-section between an antiproton and a proton by the simple geometrical
factor $4^{2/3}$.  Furthermore, $n_{\rm H}\sim 1\;{\rm cm}^{-3}$ is the number
density of Hydrogen nuclei in the Milky Way disk, $n_{\rm He}\sim 0.07 ~n_{\rm
H}$ the number density of Helium nuclei and $\sigma^{\rm ann}_{\bar p p}$ is
the proton--antiproton annihilation cross-section, which is parametrized 
by~\cite{Tan:1983de, Protheroe:1981gj}:
\begin{eqnarray}
  \sigma^{\rm ann}_{\bar p p}(T) = 
  \left\{
    \begin{array}{ll}
      661\;(1+0.0115\;T^{-0.774}-0.948\;T^{0.0151})\; {\rm mbarn}\;,
      & T < 15.5\;{\rm GeV}~, \\
      36 \;T^{-0.5}\; {\rm mbarn}\;, & T \geq 15.5\;{\rm GeV}\,. \\
  \end{array} \right. 
\end{eqnarray}

The flux of primary antiprotons from dark matter decay is highly sensitive to
the choice of propagation parameters, resulting in a variation in the flux as
large as two orders of magnitude between the MIN and MAX set of transport
parameters. This is due to the fact that theoretical determinations of
primary-to-secondary flux ratios suffer from a degeneracy between the
diffusion coefficient and the height of the magnetic diffusion zone. For
cosmic rays produced in the Galactic disk, very different sets of propagation
parameters can yield identical flux ratios. Primary fluxes from dark matter,
which are produced everywhere in the dark matter halo and not just in the
disk, vary however substantially depending on the portion of the dark matter
halo which intersects with the diffusion zone.

\begin{figure}
  \begin{center}
    \includegraphics[width=0.6\linewidth]{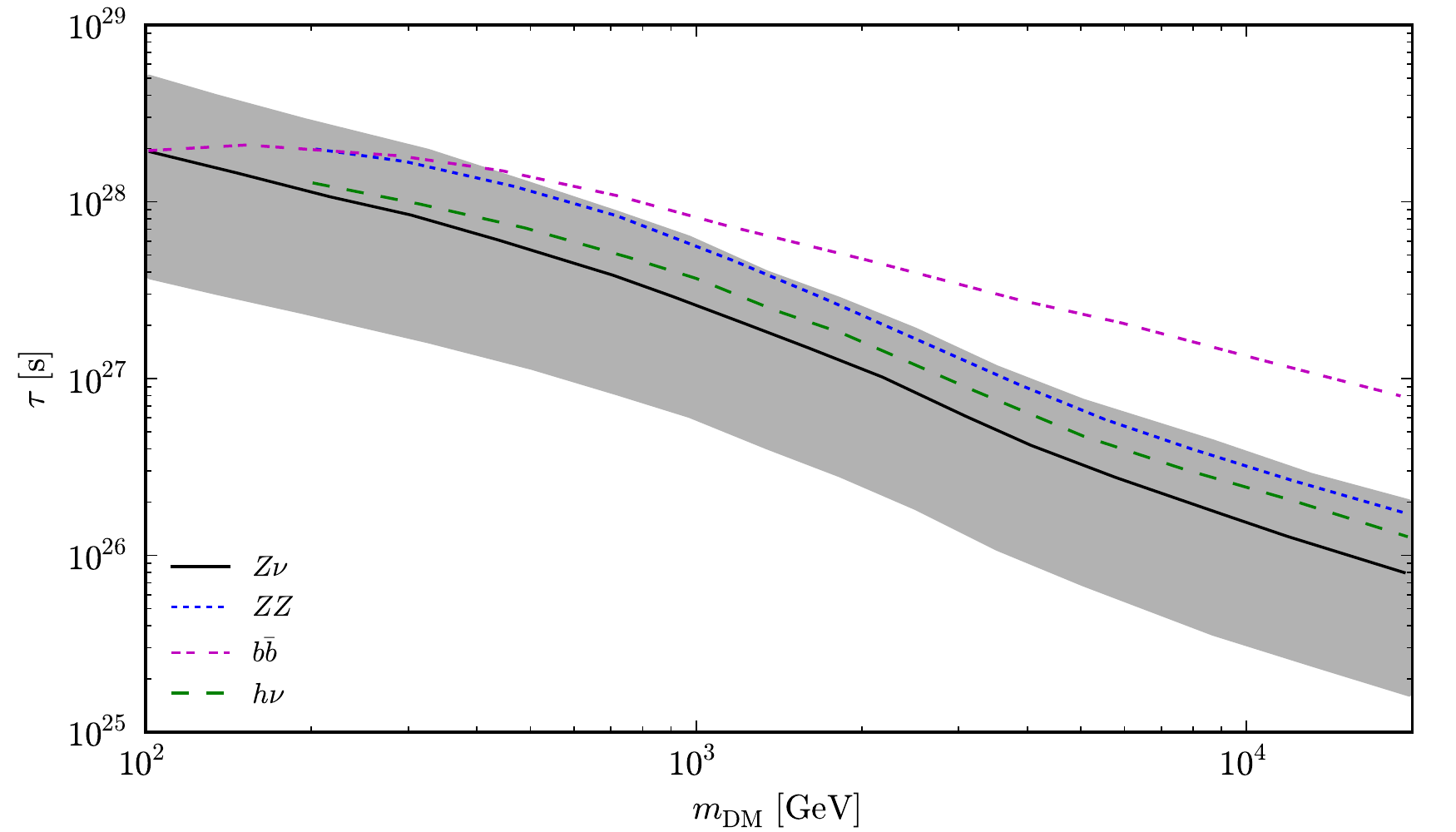}
  \end{center}
  \caption{95\% C.L. lower limits on the dark matter lifetime for decays of
    fermionic or bosonic dark matter into different final states, derived from
    \emph{antiproton measurements} by PAMELA in Ref.~\citen{Garny:2012vt}. The
    gray band indicates for the case of $Z\nu$ final states the uncertainties
    of the limits, following for the MIN/MED/MAX models as discussed in the
    text.}
  \label{fig:antiproton_limits}
\end{figure}

Antiproton constraints on decaying dark matter have been computed by a number
of authors~\cite{Cholis:2010xb,Garny:2012vt,Delahaye:2013yqa}.  Constraints
derived from the antiproton-to-proton ratio are generally more stringent than
those derived from the absolute antiproton flux. We show model-independent
constraints from Ref.~\citen{Garny:2012vt} on various dark matter decay modes
in Figs.~\ref{fig:antiproton_limits}, where the constraints are derived from
the requirement that the computed antiproton/proton ratio does not exceed the
observed values~\cite{Adriani:2010rc} at 95\% C.L. Generally, the resulting 
constraints on the dark matter lifetime are in tension with the preferred 
lifetimes for explaining the anomalous leptonic cosmic-ray measurements if 
the dark matter has significant branching ratios into hadronic final states 
(cf. Table~\ref{tab:positron_bestfit}). Improved results on the antiproton 
flux from AMS-02 can be expected in the very near future. 

\subsection{Antideuterons}
The search for antideuterons is a promising method to detect dark matter
decays due to the extremely low background fluxes expected from the spallation 
of cosmic rays on the interstellar medium in the energy range of interest for
experimental searches. In fact, all searches for cosmic antideuterons
have so far been negative and have only resulted in upper limits on the flux. 
The best present limit on the cosmic antideuteron flux was set by BESS in the range of
kinetic energy per nucleon $0.17\leq T\leq 1.15\,{\rm GeV/n}$, $\Phi_{\bar
d}<1.9\times 10^{-4} \,{\rm m}^{-2} {\rm s}^{-1} {\rm sr}^{-1} {\rm
(GeV/n)}^{-1}$~\cite{Fuke:2005it}. Interestingly, in the near future the
sensitivity of experiments to the cosmic antideuteron flux are expected to 
increase significantly, namely by more than two orders of magnitude. The Alpha Magnetic
Spectrometer (AMS-02) aboard the International Space Station
is currently searching for cosmic antideuterons in two energy windows,
$0.2\leq T\leq 0.8~{\rm GeV/n}$ and $2.2\leq T\leq 4.4~{\rm GeV/n}$, with an
expected flux sensitivity after five years $\Phi_{\bar d}=1\times 10^{-6} {\rm
m}^{-2} {\rm s}^{-1} {\rm sr}^{-1} {\rm (GeV/n)}^{-1}$ in both energy
windows~\cite{Doetinchem}. Furthermore, the balloon-borne General Antiparticle
Spectrometer (GAPS) will, starting this decade, undertake a series of
flights at high altitude over Antarctica, searching for cosmic antideuterons.
In the first phase, a long duration balloon (LDB) flight will search for
antideuterons in the range of kinetic energy per nucleon $0.1\leq T\leq
0.25~{\rm GeV/n}$ with a sensitivity $\Phi_{\bar d}=1.2\times 10^{-6} {\rm
m}^{-2} {\rm s}^{-1} {\rm sr}^{-1} {\rm (GeV/n)}^{-1}$, while in the second,
the ultra long duration balloon (ULDB) flight will search in the range
$0.1\leq T\leq 0.25~{\rm GeV/n}$ with a sensitivity $\Phi_{\bar d}=3.5\times
10^{-7} {\rm m}^{-2} {\rm s}^{-1} {\rm sr}^{-1} {\rm
(GeV/n)}^{-1}$~\cite{Doetinchem}. 

The antideuteron flux expected from spallation of cosmic rays on the
interstellar medium is peaked at a kinetic energy per nucleon $T_{\bar D}\sim
5$ GeV/n and rapidly decreases at smaller kinetic
energies~\cite{Duperray:2005si,Donato:2008yx,Donato:1999gy,Ibarra:2013qt}.  In
contrast, the spectrum of antideuterons from dark matter decays is usually
much flatter at low kinetic energies and could easily overcome the
astrophysical background for sufficiently large decay rates. For this reason,
the possibility of detecting antideuterons from dark matter decay has
received some attention over the last years~\cite{Ibarra:2009tn,Ibarra:2012cc}.

To describe the antideuteron production it is common to employ the coalescence
model~\cite{Butler:1963pp,Schwarzschild:1963zz,Csernai:1986qf,Kadastik:2009ts},
which postulates that the probability of the formation of an antideuteron out
of an antiproton-antineutron pair with given four-momenta $k_{\bar{p}}^{\mu}$
and $k_{\bar{n}}^{\mu}$ can be approximated as a narrow step function $\Theta
\left( \Delta^2+p_0^2 \right)$, where $\Delta^{\mu} = k_{\bar{p}}^{\mu} -
k_{\bar{n}}^{\mu}$ is the difference between the antiproton and antineutron
momenta. In this model, the coalescence momentum $p_0$ is the maximal relative
momentum of the two antinucleons that still allows for the formation of an
antideuteron. One can show that for $| \vec{k}_{\bar{D}}| \gg p_0$, where $
\vec{k}_{\bar{D}}=\vec k_{\bar{p}}+ \vec k_{\bar{n}}$, this ansatz leads to
the following differential antideuteron yield in momentum space:
\begin{align}
  \label{eqn:general_coal_formula}
  \gamma_{\bar{D}} \, & \frac{d^3N_{\bar{D}}}{d^3 k_{\bar{D}}} (
  \vec{k}_{\bar{D}}) = \frac18 \cdot \frac43 \pi p_0^3 \cdot  \gamma_{\bar{p}}
  \gamma_{\bar{n}} \frac{d^3 N_{\bar{p}} d^3 N_{\bar{n}}}{d^3 k_{\bar{p}} d^3
  k_{\bar{n}}} \left( \frac{\vec{k}_{\bar{D}}}{2}, \frac{\vec{k}_{\bar{D}}}{2}
  \right) \, ,
\end{align}
where the correlation between the antiproton and antineutron production in the 
hard process has been taken into account. The antideuteron yield can
then be calculated using a Monte Carlo event generator and selecting the events which
contain an antiproton--antineutron pair produced directly in the hadronization
process with a relativistic invariant momentum difference $-\Delta^2<p_0^2$
(an antiproton produced in a weak decay is separated from an antineutron
produced in the hadronization or in other weak decay by a distance much larger
than the typical range of the nuclear forces and therefore will not form a
bound state). The antideuteron yield can then be calculated in different
processes and compared to experimental data to determine the coalescence
momentum. As shown in Ref.~\citen{Ibarra:2012cc} the coalescence momentum
depends on the underlying process and on the center of mass energy. Therefore,
the coalescence momentum inferred from laboratory experiments might differ to
the actual one involved in the dark matter decay, thus introducing an
important source of uncertainty (note from
Eq.~\eqref{eqn:general_coal_formula} that the antideuteron yield scales as
$p_0^3$). This is not the case for the case of decays into weak gauge bosons,
since the antideuteron yield from $Z$ boson decay has been measured by
ALEPH~\cite{Schael:2006fd}. From this, following the procedure described
above, a coalescence momentum $p_0=192\pm 30$ MeV can be
derived.\cite{Ibarra:2012cc}

\begin{figure}
  \begin{center}
    \includegraphics[width=0.49\textwidth]{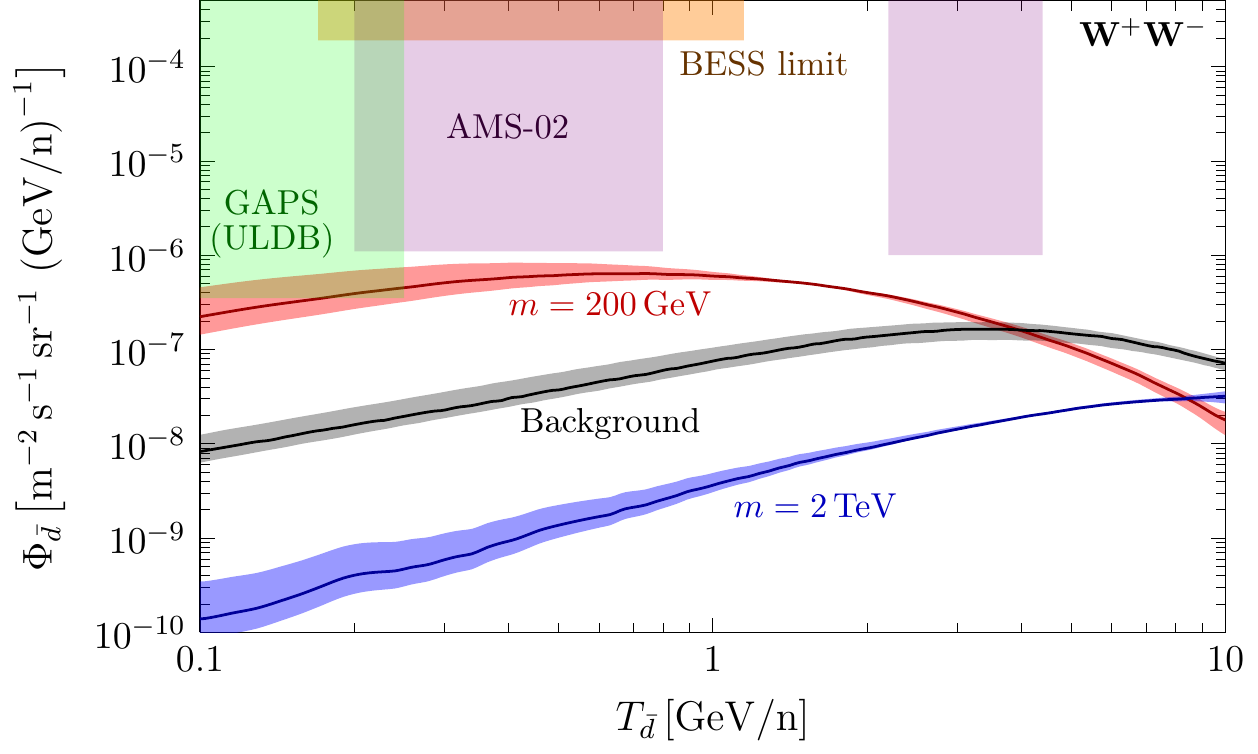}
    \includegraphics[width=0.49\textwidth]{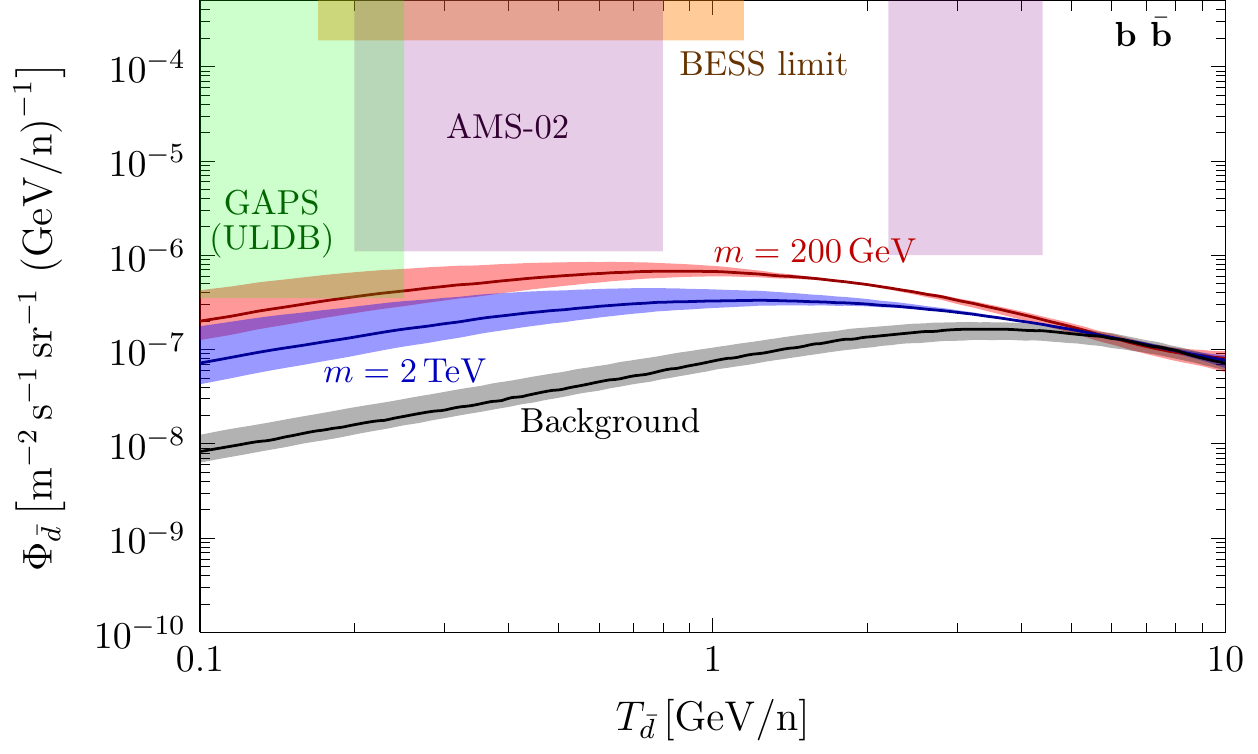}
    \caption{Maximum \emph{antideuteron} flux from dark matter decays into $W^+W^-$ (left
      plot) and $b \bar b$ (right plot) compatible with the PAMELA measurements of
      the antiproton-to-proton fraction for dark matter masses $m_{\rm DM}=200$ GeV
      (red line) and $m_{\rm DM}=2$ TeV (blue line) assuming a NFW dark matter halo
      profile, the MED propagation model and a value of the coalescence momentum
      $p_0=192$ MeV. We also show as a black line the expected background flux
      calculated in Ref.~\citen{Donato:2008yx}, also for the MED propagation model.
      The red, blue and grey shaded regions span the propagation uncertainty between
    MIN, MED and MAX parameters. Plots courtesy of Sebastian Wild.}
    \label{fig:dbar-fluxes}
  \end{center}
\end{figure}

The propagation of antideuterons in the Milky Way is analogous to the
propagation of antiprotons (see~\ref{apx:propagation} for details).  Since
antideuterons are produced by the coalescence of one antiproton and one
antineutron, it is apparent that there is a strong correlation between the
cosmic antideuteron flux and the cosmic antiproton flux. More concretely, the
non-observation of an excess in the PAMELA measurements of the cosmic
antiproton-to-proton fraction can be used to set upper limits on the decay
width in the channels producing antiprotons, which can then be translated into
upper limits on the antideuteron flux in these channels~\cite{Ibarra:2012cc}.
The corresponding upper limits on the antideuteron flux for the decays into
$W^+W^-$ and  $b \bar b$ are shown in Fig.~\ref{fig:dbar-fluxes}, for $m_{\rm
DM}=200$ GeV  and $m_{\rm DM}=2$ TeV, together with the expected background
flux calculated in Ref.~\citen{Donato:2008yx} (see also
Ref.~\citen{Ibarra:2013qt}) and the sensitivity of future and planned
experiments.  As apparent from the plot, the upper limits on the hadronic
decays set by PAMELA severely constrain the possibility of observing
antideuterons from dark matter decay at AMS-02 or GAPS, with the maximum
number of expected events being less than one at AMS-02 and one at GAPS, which would
not suffice to unequivocally attribute any possible signal to dark matter
decays at 95\% C.L.~\cite{Ibarra:2012cc}. Nevertheless, a larger number of
events could be observed, and a larger significance of the signal could be
achieved, if the upper limit on the decay width into antiprotons is reduced
and if the coalescence momentum is enhanced -- always under the assumption
that the PAMELA limits on an exotic component in the antiproton-to-proton
fraction are saturated.\cite{Fornengo:2013osa} Unfortunately, despite the
various sources of uncertainty, the observation of an antideuteron flux at
AMS-02 or GAPS from dark matter decays seems challenging.

\section{Gamma Rays}
\label{sec:gamma}
For dark matter lifetimes of the order $10^{26}$--$10^{29}$s, the
high-energetic photons potentially produced by dark matter decay could
contribute to the gamma-ray fluxes measured at Earth at an observable level.
In the case of hadronic decays, these photons would predominantly come from
$\pi^0\to\gamma\gamma$, whereas the decay into light charged leptons would,
e.g.,~give rise to intense final state radiation. Highly energetic electrons and
positrons from dark matter decay can further upscatter photons from the
interstellar radiation field (ISRF) into the gamma-ray energy range, and emit
bremsstrahlung when interacting with the interstellar medium. For
definiteness, we will focus here on gamma-ray energies above 100 MeV.

One general advantage of the gamma-ray channel is the preservation of spectral
and spatial information (in contrast, antimatter suffers energy losses and
undergoes diffusion processes). Though in the case of dark matter annihilation signals, the
characteristic signal morphology can greatly help to discriminate signals from
backgrounds, for dark matter \emph{decay} signals the most stringent
constraints come often from the \emph{isotropic} (and therefore mostly
extragalactic) signal component~\cite{Bertone:2007aw}; anisotropies play only
a subdominant role~\cite{Ibarra:2009nw}. A convincing identification of an
isotropic signal would have to rely on distinctive spectral features, such as gamma-ray
lines or pronounced bumps or cutoffs from final state radiation. Lastly,
predictions for the gamma-ray emission from dark matter decay do not suffer
from uncertainties in signal enhancement due to dark matter substructure as
in the case of annihilation signals.

We will start with a brief overview of the individual signal contributions,
which is followed by a summary of the results of current searches.

\subsection{Signals}
The most relevant contributions to the decay signal are firstly the `prompt'
photons that are directly generated during the decay, and secondly the
subsequent inverse Compton emission from the prompt electrons and positrons.
We will discuss both components separately.

\subsubsection{Prompt Radiation}
The prompt gamma-ray flux from dark matter decays can be broadly split in a
Galactic and an extragalactic component. The differential flux of signal
photons from our Galaxy is given by
\begin{equation}
  \frac{d\Phi_\text{halo}}{dE_\gamma}(\psi) =\frac{1}{4\pi}
  \sum_f \frac{\Gamma_f}{m_{\rm DM}} \frac{dN^f_\gamma}{dE_\gamma} 
  \underbrace{\int_0^\infty ds\, \rho_{\rm halo}[r(s,\psi)]}_{\equiv J}\;, 
\label{eq:flux-gamma-decay}
\end{equation}
where $\psi$ denotes the angle towards the Galactic center, $\Gamma_f$ is the
decay width corresponding to the final state $f$, and $dN^f_\gamma/dE_\gamma$
is the energy spectrum of photons produced in that channel. Furthermore,
$\rho_{\rm halo}(r)$ denotes the density distribution of dark matter particles
in our Galaxy as a function of the galactocentric distance $r(s,\psi) =
\sqrt{s^2+R^2_\odot-2 s R_\odot \cos \psi}$, with $R_\odot=8.5\,{\rm kpc}$
being the distance of the Sun to the Galactic center, and $\cos\psi=\cos b\cos
\ell$, where  $b$ and $\ell$ are the Galactic latitude and longitude, respectively. 
Commonly adopted density profiles $\rho_{\rm halo}$ were discussed in
Section~\ref{sec:Introduction}.

\begin{figure}
  \begin{center}
    \includegraphics[width=0.7\linewidth]{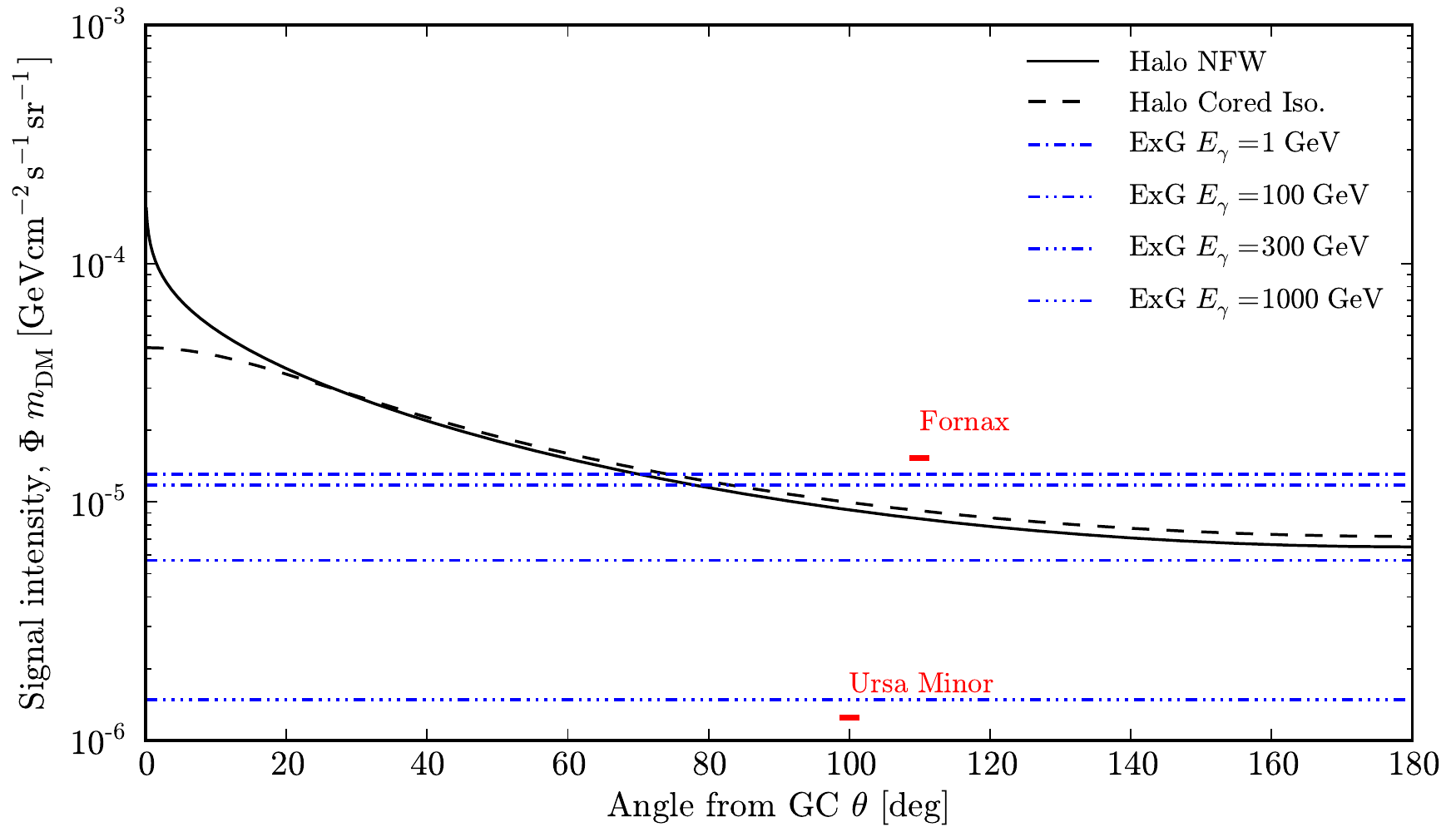}
  \end{center}
  \caption{Morphology and intensity of the dark matter decay signal. We show
    the signal intensity as function of the distance towards the Galactic
    center $\psi$, assuming a lifetime of $\tau_{\rm DM}=(m_\text{DM}/\rm GeV)^{-1}\
    10^{26}\,s$ and $\gamma\nu$ final state. 
    The solid and dashed lines correspond to the diffuse Galactic
    contribution, the dash-dotted line to the extragalactic signal which
    depends on the emission energy $E_\gamma$. For comparison,
    the signal intensities predicted for the Fornax cluster and the dwarf
    spheroidal Ursa Minor, both averaged over a region of $1^\circ$ radius, are
    shown by the red bars.}
  \label{fig:gamma-morphology}
\end{figure}

For the case of a NFW (cored isothermal) profile, the predicted dark matter signal flux
from the Galactic halo as function of $\psi$ is shown by the solid (dashed)
line in Fig.~\ref{fig:gamma-morphology}. For definiteness, we assume a dark
matter
lifetime of $\tau_{\rm DM}=(m_\text{DM}/\rm GeV)^{-1}\ 10^{26}\,s $, and $100\%$
annihilation into a $\gamma\nu$ final state. The difference between the two
profiles becomes substantial at angles $\psi\lesssim 20^\circ$ close to the
Galactic center, and leads to signal predictions that differ by a factor of
four and more at $\psi\lesssim 0.1^\circ$. At larger values of $\psi$ the main
uncertainty is the overall normalization of $\rho_{halo}$.  \medskip

The extragalactic contribution to the gamma-ray signal is generated by the
decay of dark matter particles at cosmological distances. It is largely
isotropic and affected by the redshift as well as as the finite optical depth
of the Universe. The differential flux is given by
\begin{equation}
    \frac{d\Phi_\text{eg}}{dE_\gamma} =\frac{1}{4\pi}\frac{\Omega_{\rm DM}
    \rho_\text{c} }{m_{\rm DM}}
    \int_0^\infty dz \sum_f \Gamma_f \frac{1}{H(z)}
    \frac{dN^f_\gamma}{dE_\gamma}\left[(z+1)E_\gamma\right]
    \;e^{-\tau(E_\gamma,z)}\;.
\label{eq:gamma-dec-EG}
\end{equation}
Here, $\rho_\text{c}=4.9\times10^{-6}\GeV\cm^{-3}$ denotes the critical
density of the Universe, and $H(z)=H_0
\sqrt{\Omega_\Lambda+\Omega_\text{m}(z+1)^3}$ is the Hubble rate as a function
of redshift $z$.\footnote{We adopt the parameters $\Omega_\Lambda=0.69$,
$\Omega_\text{m}=0.31$, $\Omega_\text{DM}=0.26$ and $h\equiv
H_0/100\,\text{km}\;\text{s}^{-1}\,\text{Mpc}^{-1}=0.68$, as derived from
Planck+WP+highL+BAO data (see Ref.~\citen{Ade:2013lta}).} With the factor
$e^{-\tau}$, we incorporate attenuation effects due to pair-production and --
in the TeV regime -- photon-photon scattering on the intergalactic background
light (IBL).  The attenuation factor is determined by the optical depth
$\tau(E_\gamma,z)$, for which we adopt the results from
Ref.~\citen{Cirelli:2010xx}. These results are based on the AEGIS multi-wavelength
analysis presented in Ref.~\citen{Dominguez:2010bv}, and compatible with
recent HESS~\cite{Aliu:2008ay} and Fermi~\cite{Abdo:2010kz} observations.

In Fig.~\ref{fig:gamma-morphology} the horizontal lines show the isotropic
flux of extragalactic gamma rays for different energies $E_\gamma$ of the
prompt photons (as above, we integrate over all gamma-ray energies to
calculate the flux). For our reference Galactic dark matter halo and photon
energies below about 100 GeV, the extragalactic dominates the Galactic signal
at angles $\psi\gtrsim90^\circ$. At these energies, half of the extragalactic
signal stems from redshifts $z\gtrsim0.8$. However, at higher energies the
attenuation effects start to suppress the extragalactic signal, and at
$\gtrsim800$~TeV energies only the local Universe ($z\lesssim0.1$) remains
observable in gamma rays.  \medskip

Besides the Galactic and cosmological components, massive nearby dark matter
halos -- such as galaxy clusters or Milky Way satellite galaxies -- contribute to
the overall signal (cf.~Ref.~\citen{Cuesta:2010ex}).  They appear as
point-like or marginally extended objects and could be used to identify a
signal above the diffuse background.  The corresponding signal flux can be
calculated as shown in Eq.~\eqref{eq:flux-gamma-decay}, with the
substitutions $\rho_{\rm halo}\to \rho_{\rm src}$ and $R_\odot\to D$;
$\rho_{\rm src}$ is the dark matter distribution of the target and $D$ its
distance to the Earth.  In Fig.~\ref{fig:gamma-morphology}, we show for two
interesting targets -- the dwarf spheroidal Ursa Minor and the galaxy cluster
Fornax -- the associated fluxes averaged over a region with a radius of
$1^\circ$ (we adopt here values from Ref.~\citen{Dugger:2010ys}). The
predicted signal from these and other targets is at most marginally more
intense (and in most cases weaker) than the diffuse Galactic and extragalactic
signal components.

\subsubsection{Inverse Compton Scattering}
High-energetic electrons and positrons that are produced during the decay of
dark matter particles can upscatter photons from the interstellar radiation
field (ISRF) to gamma-ray energies via inverse Compton
scattering~\cite{Blumenthal:1970gc}. Besides the CMB, the ISRF includes thermal
dust radiation as well as diffuse starlight~\cite{Porter:2005qx}. We will here briefly
review the calculation of the Galactic ICS signal, and refer to 
Refs.~\citen{Ishiwata:2009dk, Ibarra:2009dr} for the extragalactic counterpart.

The differential production rate of gamma rays with energies around $E_\gamma$
at position ${\vec r}$ by the inverse Compton scattering of highly energetic
electrons and positrons on photons of the ISRF reads
\begin{equation}
  \begin{split}
    &\frac{dR^{\rm IC}_\gamma(\vec r)}{d E_\gamma}=\int_0^\infty d\epsilon
    \int_{m_e}^{\infty} dE_e\; \frac{d\sigma^{\rm
    IC}(E_e,\epsilon)}{dE_\gamma} f_{e^\pm}(E_e,{\vec r})
    f_{\rm ISRF}(\epsilon,\vec{r})\;.
  \end{split}
  \label{eqn:IC-rate}
\end{equation}
Here, the differential number density of electrons (or positrons) from dark
matter decay is given by $f_{e^\pm}(E_e,{\vec r})$, and the differential
number density of the ISRF is $f_{\rm ISRF}(\epsilon,\vec{r})$.  Furthermore,
$d\sigma^\text{IC}/dE_\gamma(E_e, \epsilon)$ is the differential ICS cross
section for an electron with energy $E_e$ to up-scatter ISRF photons from
$\epsilon$ to $E_\gamma$. It follows from the Klein-Nishina formula and is
given by
\begin{equation}
  \begin{split}
    &\frac{d\sigma^{\rm IC}(E_e,\epsilon)}{dE_\gamma}=
    \frac{3}{4}\frac{\sigma_{\rm T}}{\gamma_e^2\, \epsilon} \left[2q\ln q + 1
    + q - 2q^2
      +\frac{1}{2}\frac{(q\Gamma)^2}{1+q\Gamma}(1-q)\right],
  \end{split}
  \label{eqn:ICrate}
\end{equation}
where $\sigma_{\rm T}=0.67\,{\rm barn}$ denotes the Compton scattering cross
section in the Thomson limit, $\gamma_e\equiv E_e/m_e$ is the Lorentz factor
of the electron, $m_e=511\,{\rm keV}$ is the electron mass, and we defined the
quantities $\Gamma\equiv 4\gamma_e\epsilon/m_e$ and $q\equiv
E_\gamma/\Gamma(E_e-E_\gamma)$. Eq.~\eqref{eqn:ICrate} holds in the limit
where $\epsilon,m_e\ll E_e$, and kinematics and the neglect of down-scattering
require that $\epsilon\leq E_\gamma \leq (1/E_e + 1/4\gamma_e^2 \epsilon)^{-1}
\equiv E_\gamma^\text{max}$. In case of CMB photons with
$\epsilon\sim2.4\times10^{-4}\eV$, ICS photons are produced up to energies of
$E_\gamma\sim3.7\GeV\ E_{{\rm TeV}, e}^2$, with electron energies given in
units of TeV. 

The number density of electrons and positrons from dark matter decays,
$f_{e^\pm}(E_e,\vec r)$, follows from solving the full transport equation
Eq.(\ref{eqn:transport}). At energies above a few tens of GeV, energy losses
dominate, which allows some approximate, simple analytical solutions (see
e.g.~discussion in Ref.~\citen{Ibarra:2009nw}). A commonly adopted model for
the ISRF is the one in Ref.~\citen{Porter:2005qx}. Using these distributions
and Eq.~\eqref{eqn:IC-rate}, the gamma-ray flux from ICS that is received at
Earth as function of Galactic longitude $\ell$ and latitude $b$ is given by
\begin{equation}
  \frac{d\Phi_\text{halo,IC}}{dE_\gamma}(\ell,b) =
  2\cdot\frac{1}{4\pi} 
  \int_0^\infty ds\;\frac{dR^{\rm IC}_\gamma[r(s,\ell,b)]}{d
  E_\gamma}\;,
  \label{eqn:IC-flux-halo}
\end{equation}
where the factor of 2 takes into account the fact that both dark matter
electrons and positrons contribute equally to the total flux of gamma rays.

Starlight is brightest towards the Galactic center; together with the geometry
of the Galactic cosmic-ray diffusion zone this causes the ICS dark matter signal to be
elongated and aligned with the Galactic plane -- in contrast to the
approximately spherical prompt component of the signal, making it potentially difficult
to discriminate from an ICS signal from dark matter
annihilation~\cite{Boehm:2010qt}. However, due to the large Galactic
foregrounds in direction of the Galactic center it is more efficient to search
for the signal at the Galactic poles. There, the predicted signal intensity
depends critically on the electrons and positrons that are generated by dark
matter decay inside the Galactic halo, but \emph{outside} of the diffusion
zone.

Search strategies for dark matter signals often build upon specific morphological and
spectral signal characteristics and depend on the dark matter model at hand.
For decaying dark matter with masses in the GeV--TeV regime, the most relevant
targets are the isotropic gamma-ray background (IGBG), the Galactic halo,
galaxy clusters, and sharp spectral signatures.

\subsubsection{Morphological Features}
As shown in Fig.~\ref{fig:gamma-morphology}, at angles of a few tens of degrees
with respect to the Galactic center, the dark matter decay signal is most intense and the
vast majority of signal photons would come from the Galactic halo itself.
Unfortunately, the Galactic center is also the region with the largest
foregrounds at GeV energies.  Thus, for decaying dark matter signals, the best
regions of interest lie offset from the Galactic Center, 
above or below the Galactic
plane~\cite{Bertone:2007aw, Ibarra:2009nw}.  In fact, it turns out that
measurements of the IGBG, which dominates the diffuse emission observed in
direction of the Galactic poles, provide some of the most stringent
constraints on dark matter decay.

\begin{figure}
  \begin{center}
    \includegraphics[width=0.6\linewidth]{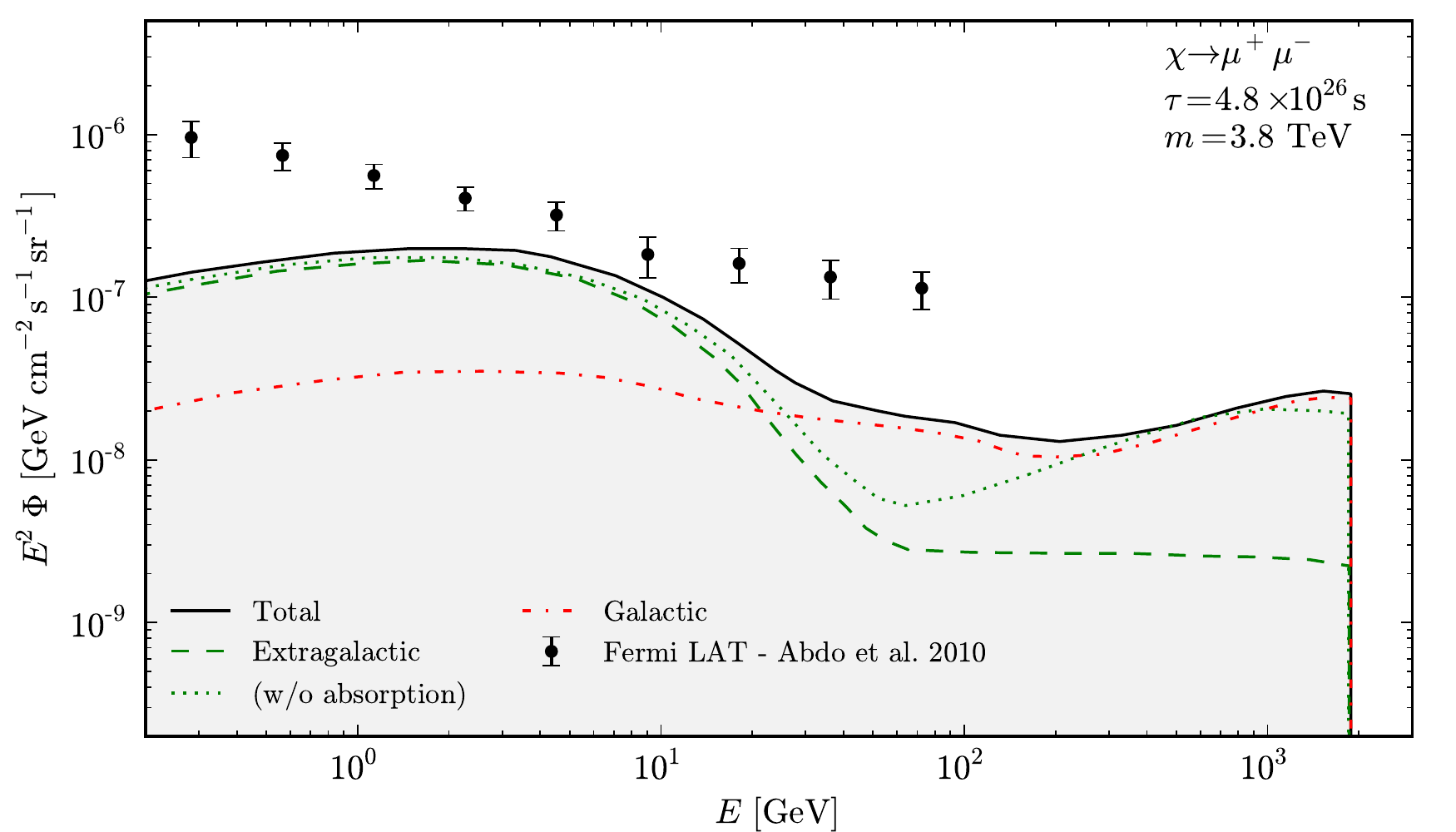}
  \end{center}
  \caption{The extragalactic gamma-ray background as measured by the Fermi
    LAT~\cite{Abdo:2010nz, ackermann:fermisymp2012}, compared to exemplary
    predictions for decay into $\mu^+\mu^-$ final states. The dashed (dotted)
    line shows the extragalactic contribution with (without) absorption
    effects included; the dash-dotted line is the Galactic contribution.
    Fluxes are taken from Ref.~\citen{Cirelli:2012ut} and correspond to the flux
    predictions for the direction $\ell=180^\circ$ and $b=0^\circ$ (see text
    details);  ICS emission dominates at energies below $\sim60\GeV$.}
  \label{fig:gamma-signal}
\end{figure}

Technically, the IGBG is the -- by construction -- isotropic flux that remains
after subtracting the Galactic diffuse emission and known point sources.  The
IGBG was recently determined by Fermi LAT up to energies of 100
GeV~\cite{Abdo:2010nz}, preliminary results up to 400 and 580~GeV were
presented in Refs.~\citen{Ackermann:TeVPA2011, ackermann:fermisymp2012}. The
measured fluxes are shown in Fig.~\ref{fig:gamma-signal}, together with an
exemplary dark matter signal. The IGBG is usually attributed to gamma rays
originating from outside of our Galaxy, and thought to be made up entirely by
unresolved point-like sources, like blazars, millisecond pulsars or star-
forming galaxies (see Ref.~\citen{Calore:2013yia} and references therein).

As discussed above, dark matter decay at cosmological distances would generate
a nearly isotropic gamma-ray signal, which would entirely contribute to the
IGBG. However, close to the Galactic poles the angular dependence of the
Galactic signal is in case of dark matter decay relatively weak, such that
this part of the Galactic signal would be likely misidentified as being
extragalactic. As a very conservative lower limit on the dark matter signal flux
contribution to the IGBG, one can adopt the signal strength in direction of
the anti-center, namely $\Phi_\text{eg} + \Phi_\text{halo}(\psi=180^\circ$).

\subsection{Targets and Searches}
\begin{figure}
  \begin{center}
    \includegraphics[width=1.\linewidth]{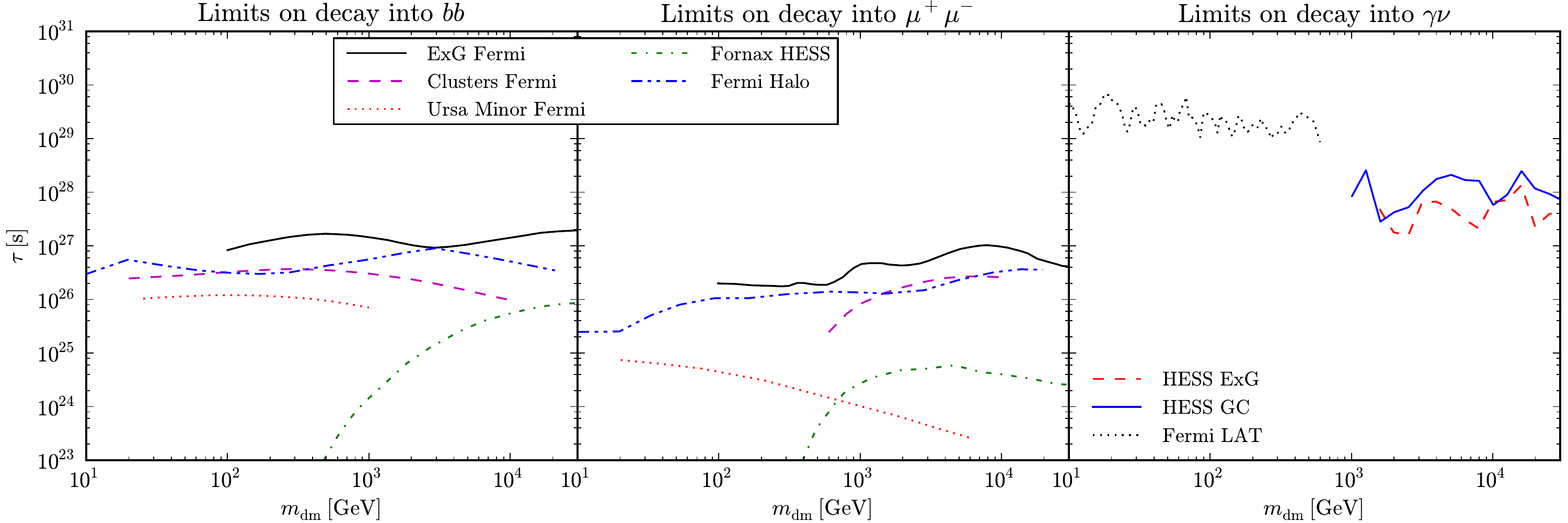}
  \end{center}
  \caption{Summary of 95\%CL lower limits on the dark matter lifetime, derived
    from gamma-ray observations, assuming different final states: \emph{left
    panel:} $\psi\to \bar bb$, \emph{central panel:} $\psi\to \mu^+\mu^-$ and
    \emph{right panel:} $\psi\to \nu\gamma$. Limits are taken from
    Refs.~\citen{Dugger:2010ys, Huang:2011xr, Cirelli:2012ut,
    Ackermann:2012rg} in case of $\bar bb$ and $\mu^+\mu^-$ final states, and
    from Refs.~\citen{Fermi-LAT:2013uma, Abramowski:2013ax} in case of
    gamma-ray lines.}
  \label{fig:gamma-limits}
\end{figure}

For a decaying dark matter signal, a simple but very robust constraint can be
derived from the requirement that the dark matter signal should not significantly 
exceed the measured IGBG; this approach was used by many groups~\cite{Chen:2009uq, Cirelli:2009dv,
Papucci:2009gd, Zhang:2009ut, Ibarra:2009dr, Ishiwata:2010am, Hutsi:2010ai,
Cirelli:2012ut}. More realistic constraints can be obtained by a subtraction of
astrophysical contributions~\cite{Calore:2011bt, Calore:2013yia}, or by a
detailed spectral analysis~\cite{Abdo:2010dk, Cirelli:2012ut}. The results
from Ref.~\citen{Cirelli:2012ut} -- based on a spectral fit to the IGBG with
the background modeled as power-law -- are shown in
Fig.~\ref{fig:gamma-limits}. They are stronger than the limits obtained by the
`robust' method by a factor of up to five~\cite{Cirelli:2012ut}.

Although, as outlined above, the monopole of the dark matter decay signal
turns out to be the most effective observable, the dipole anisotropy
towards the Galactic center can also have important effects [we define the dipole
  anisotropy here as $A=(J_{|\ell|\leq90^\circ} -
J_{|\ell|>90^\circ})/(J_{|\ell|\leq90^\circ} + J_{|\ell|>90^\circ})$].  At
higher latitudes (masking $|b|<10^\circ$), this anisotropy is of the order
$A=20\%$--$36\%$~\cite{Ibarra:2009nw}, depending on the profile of the
Galactic dark matter halo and the gamma-ray energy. It can potentially be
useful to finally discriminate a decay signal from the (truly isotropic)
astrophysical extragalactic gamma-ray background, since the observation of
spectral features in the dipole anisotropy as function of energy could
indicate a dark matter signal that dominates over a short energy
range.~\cite{Ibarra:2009nw}. Unfortunately, the recently discovered Fermi
Bubbles~\cite{Dobler:2009xz, Su:2010qj} contribute significantly to the dipole
anisotropy at higher latitudes, which complicates the extraction of a clear
dark matter induced dipole signal.\medskip

In regions closer to the Galactic Center 
(which might be actually favorable in case of ICS
signals~\cite{Zhang:2009ut}), an analysis of the diffuse gamma-ray emission
requires a careful modeling of the Galactic diffuse and point-source emission.
Conservative upper limits on the decay width can always be obtained by
requiring that the observed fluxes are not exceeded
significantly~\cite{Papucci:2009gd, Cirelli:2009dv}. Limits that result from 
attempting a subtraction of Galactic emission have been derived in
Ref.~\citen{Zhang:2009ut} using a simple reference model for the Galactic
emission, and in Ref.~\citen{Ackermann:2012rg} employing a detailed fit to the
gamma-ray data. The region of interest used in Ref.~\citen{Ackermann:2012rg},
$5^\circ<|b|<15^\circ$ and $|\ell|<80^\circ$, spans a large part of the sky
above and below the Galactic disk.  The gamma-ray flux from this region was
fitted with a model for the Galactic emission while marginalizing over a large
number of parameters that describe amongst others the details of CR diffusion
and source distributions, the Galactic magnetic field and uncertainties in the
interstellar medium. In the energy range 0.1--300~GeV no residual emission was
found after subtraction of the best-fit model. The resulting limits on an
additional dark matter signal are quite competitive and shown in
Fig.~\ref{fig:gamma-limits} as dash-dot-dotted lines ($3\sigma$ limits).  In
case of $\bar bb$ final states, they are comparable to the results obtained
from Galaxy cluster observations, which will be discussed next. For
$\mu^+\mu^-$ final states, starlight leads to a strong ICS signal even for
dark matter masses considerably below $1\TeV$, where the cluster limits become
very poor.
\medskip

Nearby galaxy clusters that are interesting for dark matter searches have a typical
angular scale of $\mathcal{O}(1^\circ)$ in the sky, with a signal profile that
increases steeply towards their center and with peak intensities much higher
than the Galactic signal.  However, when taking into account finite angular
resolution and statistics of instruments like the Fermi LAT, the maximum
signal intensities are comparable to or smaller than the Galactic halo signal.
This is shown in Fig.~\ref{fig:gamma-morphology} for the Fornax clusters,
which is one of the brightest targets (intensities are averaged
over a region of $1^\circ$ radius).

The gamma-ray signal from dark matter decaying in nearby galaxy clusters would
be extended enough to be seen as an marginally extended source by Fermi
LAT. Refs.~\citen{Dugger:2010ys, Huang:2011xr} searched for such emission in
up to eight
clusters using Fermi LAT data (see also Refs.~\citen{Bertone:2007aw,
Garny:2010eg, Ke:2011xw, Luo:2011bn,
Combet:2012tt, Cirelli:2012ut}). No signal was found, and upper limits were
derived on the decay rate by merging the likelihood functions of the different
observations into a combined likelihood that also takes into account cluster
mass uncertainties associated with the underlying X-ray
measurements~\cite{Huang:2011xr}.  For the
cases of decay into $\bar b b$ and $\mu^+\mu^-$ final states, we show the
limits in Fig.~\ref{fig:gamma-limits}. Note that conservatively only ICS on
the CMB has been taken into account when calculating the limits.  They turn
out to be weaker than the limits obtained from the IGBG, but are still strong
enough to be in mild tension with the decaying dark matter explanation of the
positron excess observed by PAMELA and AMS-02 (see discussion above).  For comparison, we
also show the limits that were derived from HESS observations of the Fornax
cluster.  In the energy range of interest, these limits are hardly
competitive~\cite{Cirelli:2012ut}.  \medskip

Other targets considered in the literature include dwarf spheroidal galaxies
(as illustrative example Ursa Minor is shown in
Fig.~\ref{fig:gamma-morphology})~\cite{Bertone:2007aw, Essig:2009jx,
PalomaresRuiz:2010pn, Dugger:2010ys} and the angular power
spectrum~\cite{Ibarra:2009nw}, but the corresponding constraints are
significantly weaker than the ones derived from the IGBG.

\subsubsection{Spectral Features} 
As discussed above, the most constraining gamma-ray observable for decaying
dark matter at GeV--TeV energies is the IGBG. One clear disadvantage of the
IGBG is the absence of any strong morphological feature that could help to
discriminate a dark matter signal from other astrophysical sources (though the
dipole-anisotropy due to our off-center position in the halo might 
help~\cite{Bertone:2007aw, Ibarra:2009nw}). However,
the IGBG can be measured in a very large fraction of the sky away from the
Galactic disc. As a consequence, in contrast to many other observables, the
statistical error of the flux is typically extremely small. This turns the
IGBG into an
excellent target to search for specific \emph{spectral} signatures as smoking
gun signals for dark matter decay.

The most prominent spectral signature would be a gamma-ray line from two-body
decays into photons or photon pairs. If both final state particles are
approximately massless, as e.g.~in $\psi\to\gamma\nu$, the gamma-ray line is 
observed at an energy of half the dark matter mass, $E_\gamma=m_{\rm DM}/2$. 
Such gamma-ray lines are often present as one-loop corrections to the main decay
channel~\cite{Garny:2010eg}, and typically significantly suppressed.  Among the
theoretical decaying dark matter scenarios that actually predict the emission
of \emph{strong} lines are gravitino dark matter with a mild violation of
$R$-parity~\cite{Ibarra:2007wg} or hidden sector dark matter decaying via
higher-dimensional operators~\cite{Arina:2009uq}. We will discuss these models
below in Section~\ref{sec:models}.

A closely related spectral signature are the box-shaped spectra that are
produced in cascade decays like $\psi\to\phi\phi$, with a subsequent $\phi \to
\gamma\gamma$~\cite{Ibarra:2012dw}. The corresponding gamma-ray spectrum has
step-like edges, $dN_\gamma/dE\propto \theta(E^+-E)\theta(E-E^-)$, where the
$E^\pm$ are functions of $m_{\rm DM}$ and $m_\phi$~\cite{Ibarra:2012dw}.  In
the limit $m_\phi\to m_{\rm DM}$, the signal approaches a monochromatic line
at an energy $E_\gamma=m_{\rm DM}/4$.

Different groups have explicitly searched for these signatures, using Fermi
LAT~\cite{Abdo:2010nc, Vertongen:2011mu, Bringmann:2012vr, Weniger:2012tx,
Ackermann:2012qk, Fermi-LAT:2013uma} and HESS~\cite{Garny:2010eg,
Abramowski:2013ax} data. No indication for a signature that could be
attributed to decaying dark matter was found. (The 130~GeV feature at the
Galactic center tentatively observed by the Fermi LAT~\cite{Bringmann:2012vr,
Weniger:2012tx, Tempel:2012ey, Su:2012ft, Fermi-LAT:2013uma} can be
interpreted as a signal from dark matter decay~\cite{Buchmuller:2012rc,
Kyae:2012vi}, but its spatial profile strongly disfavors this
possibility~\cite{Bringmann:2012ez}.)

At gamma-ray energies 5--300 GeV, the strongest limits come from
Fermi LAT~\cite{Fermi-LAT:2013uma} (limits down to 1~GeV were derived in
Ref.~\citen{Vertongen:2011mu}). These limits are based on nearly all-sky
observations, excluding only the Galactic discs with its large foregrounds. As
shown in Fig.~\ref{fig:gamma-limits}, they are stronger than limits on purely
hadronic and leptonic final states by two to three orders of magnitude.  Above
500 GeV and up to 25 TeV, the best limits on gamma-ray lines come from HESS.
Ref.~\citen{Abramowski:2013ax} presented flux upper limits from a spectral
analysis of the IGBG as well as the Galactic center. We translate these limits
into upper limits on the branching width into $\gamma\nu$ final states, see
Fig.~\ref{fig:gamma-limits} (right panel). They essentially extend the strong
results from Fermi LAT to higher energies, though at a somewhat lower level.
Fermi LAT limits on box-shaped spectra were derived
Ref.~\citen{Ibarra:2012dw}. Next-generation instruments like CTA and GAMMA-400
are expected to improve the limits on gamma-ray lines by up to an order of
magnitude within the upcoming ten years or so~\cite{Bergstrom:2012vd}.


\section{Neutrinos}
\label{sec:neutrinos}

The calculation of the neutrino flux from dark matter decays proceeds along
similar lines as for gamma-rays, with the fluxes being analogous to those given by
Eqs.(\ref{eq:flux-gamma-decay},\ref{eq:gamma-dec-EG}). Namely,
\begin{equation}
  \frac{d\Phi_\text{halo}}{dE_\nu}(\psi) =\frac{1}{4\pi}
  \sum_f \frac{\Gamma_f}{m_{\rm DM}} \frac{dN^f_\nu}{dE_\nu} 
  \int_0^\infty\rho_{\rm halo}[r(s,\psi)] ds
\label{eq:flux-neutrino-decay}
\end{equation}
for the decay of dark matter particles in the Milky Way halo, and
\begin{equation}
    \frac{d\Phi_\text{eg}}{dE_\nu} =\frac{1}{4\pi}\frac{\Omega_{\rm DM}
    \rho_\text{c} }{m_{\rm DM}}
    \int_0^\infty dz \frac{1}{H(z)}  \sum_f \Gamma_f
    \frac{dN^f_\nu}{dE_\nu}\left[(z+1)E_\nu\right]
\label{eq:neutrino-dec-EG}
\end{equation}
for the redshifted contribution from dark matter decays at cosmological
distances. Note that for GeV-TeV dark matter masses, and in contrast to
gamma-rays, neutrinos are not significantly absorbed while they propagate to
the Earth.

After being produced in the decay of dark matter particles, neutrinos undergo
flavor oscillations. Neglecting CP violating effects during propagation and
taking the best fit values of the neutrino oscillation parameters
$\sin^2\theta_{12}=0.30$, $\sin^2\theta_{23}=0.41$,
$\sin^2\theta_{13}=0.023$~\cite{GonzalezGarcia:2012sz}, the conversion
probabilities read:
\begin{equation}
  \begin{split}
    &P(\nu_e\leftrightarrow\nu_e) =0.56\,, P(\nu_e\leftrightarrow\nu_{\mu})=0.28\;, \\
    &P(\nu_e\leftrightarrow\nu_{\tau}) =0.16\,,
    P(\nu_{\mu}\leftrightarrow\nu_{\mu})=0.34\;,\\
    & P(\nu_{\mu}\leftrightarrow\nu_{\tau})=0.37, P(\nu_{\tau}\leftrightarrow\nu_{\tau}) =0.46\;.
  \end{split}
  \label{oscprob}
\end{equation}
Thus, a primary neutrino flux in a specific flavor is redistributed almost
equally into all neutrino flavors during propagation and any flavor
information is lost.

The neutrino spectrum from dark matter decay depends on the concrete decay
channel. The simplest possibility for dark matter decay, and which arises in
many well motivated models, is the direct decay into two neutrinos, for a
scalar particle, or into $\gamma \nu$, for a fermion. In this case,
the resulting spectrum consists of a monochromatic line and an integral 
of the redshifted line from the extragalactic signal. In the case of decays
into $(Z^0, h)\nu$, in addition to this signal there is also a continuum
of neutrinos produced in the decay and fragmentation of the $Z^0$ and Higgs bosons.  
Another simple channel is the three body decay into electrons, $\psi\rightarrow
e^+e^-\nu$ which in the most common scenarios (namely when the decay is
mediated by a heavy scalar or a heavy vector boson) has the familiar
triangular shape when plotted in a logarithmic axis. Lastly, decay modes into
other particles particles which eventually produce neutrinos in their decay or
fragmentation (such as muons, taus, or weak gauge bosons) generate a softer
neutrino energy spectrum.

The most important background for the detection of a neutrino flux from dark
matter decays comes from neutrinos produced in cosmic-ray interactions with
the Earth's atmosphere, and which have a flux that was calculated in
Ref.~\citen{Honda:2006qj}. Other sources of background are tau
neutrinos from the decay of charmed particles that are also produced in
cosmic-ray collisions with the atmosphere~\cite{Pasquali:1998xf},  neutrinos
produced in cosmic-ray interactions with the solar
corona~\cite{Ingelman:1996mj} and neutrinos produced in cosmic-ray
interactions with the interstellar medium in the Milky
Way~\cite{Athar:2004um}. 

\begin{figure}
 \centering
 \includegraphics[width=0.48\linewidth]{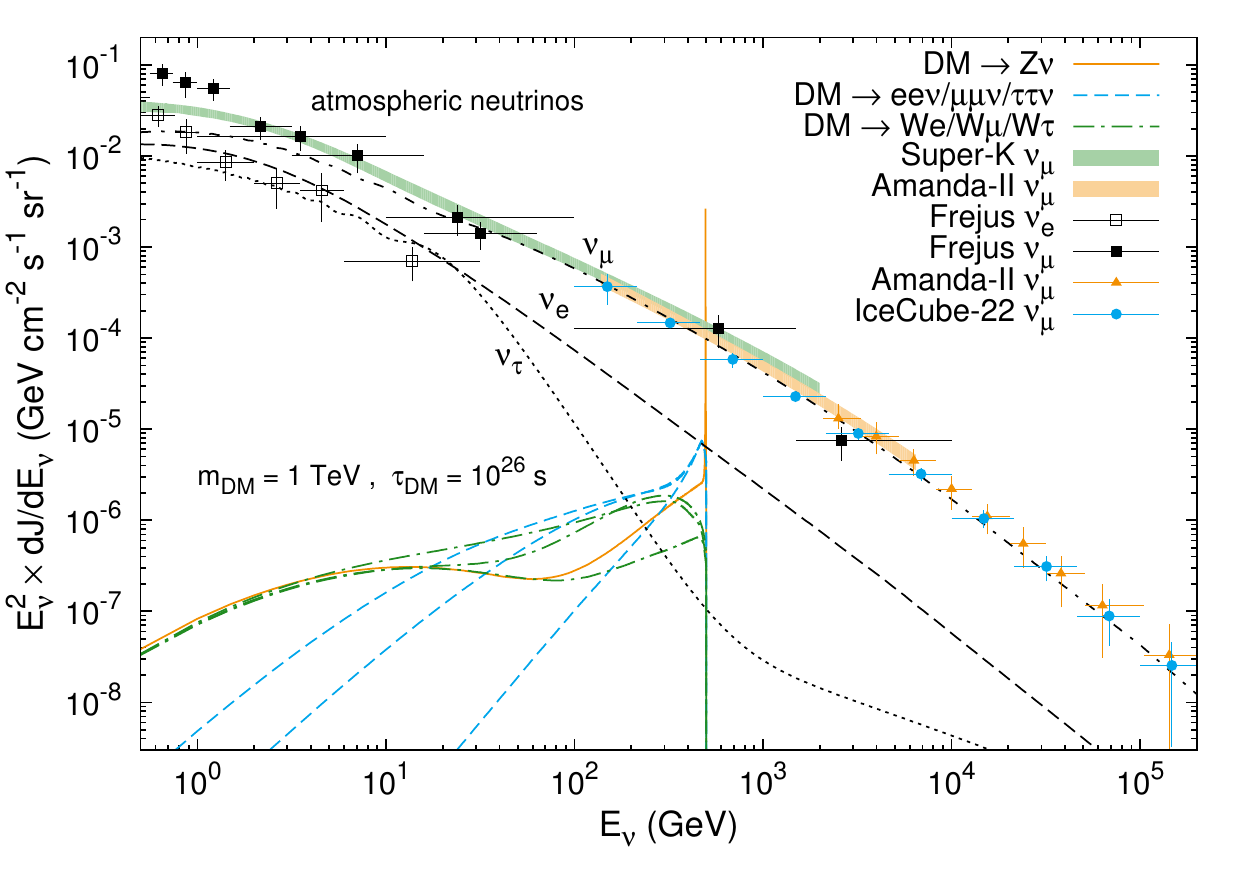}
 \includegraphics[width=0.48\linewidth]{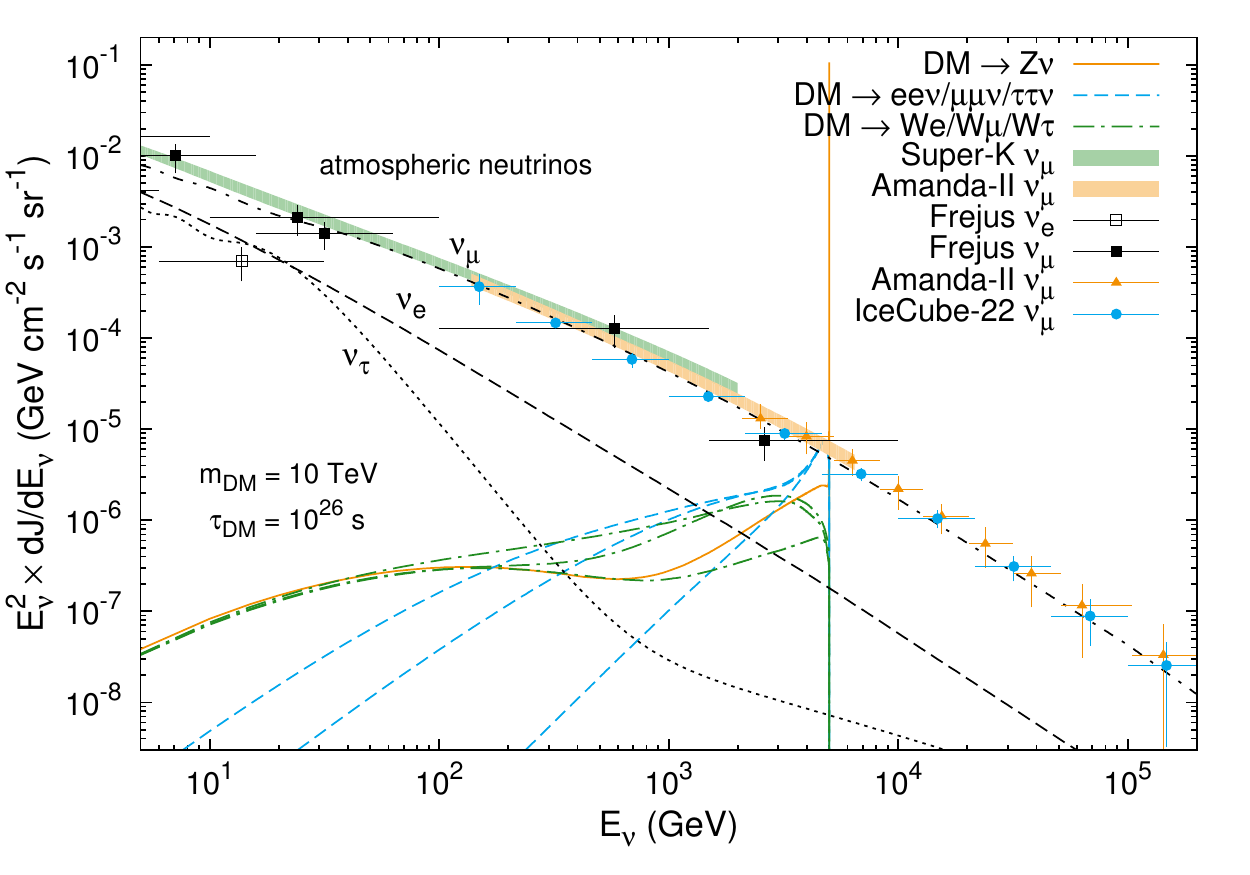}
 \caption{Neutrino spectra for different decay channels of a fermionic dark
 matter candidate. From Ref.~\citen{Covi:2009xn}. See text for details.}
 \label{fig:neutrino-spectra-fermion}
\end{figure}

The neutrino spectra at the Earth from various decay modes are shown for the
case of a fermionic dark matter particle in
Fig.~\ref{fig:neutrino-spectra-fermion}, fixing for concreteness the dark
matter mass to 1 TeV or 10 TeV and the lifetime to $\tau_{\rm DM}= 10^{26}\s$, which is
the reference value quoted in Eq.~\eqref{eq:target}. The spectra include both
the contribution from dark matter decays in the halo as well as the redshifted
contribution from decays at cosmological distances. The plot also shows  the
expected atmospheric background and the data measured by the
Fr\'ejus~\cite{Daum:1994bf}, Super-Kamiokande~\cite{GonzalezGarcia:2006ay},
AMANDA-II~\cite{Abbasi:2009nfa} and
IceCube~\cite{chirkin:ICRC2009,Abbasi:2011jx} experiments.  As apparent form
the plot, the predicted flux from dark matter decay lies considerably below
the measured muon neutrino flux. Therefore, detecting a dark matter signal
requires an efficient suppression of the backgrounds. As argued in
Ref.~\citen{Covi:2009xn} this could be achieved by considering neutrinos
arriving from all directions in the sky, since this choice optimizes the
significance of the signal, and  by exploiting the spectral information
carried by the neutrinos produced in the dark matter decay.

\begin{figure}
  \centering
  \includegraphics[width=0.60\linewidth]{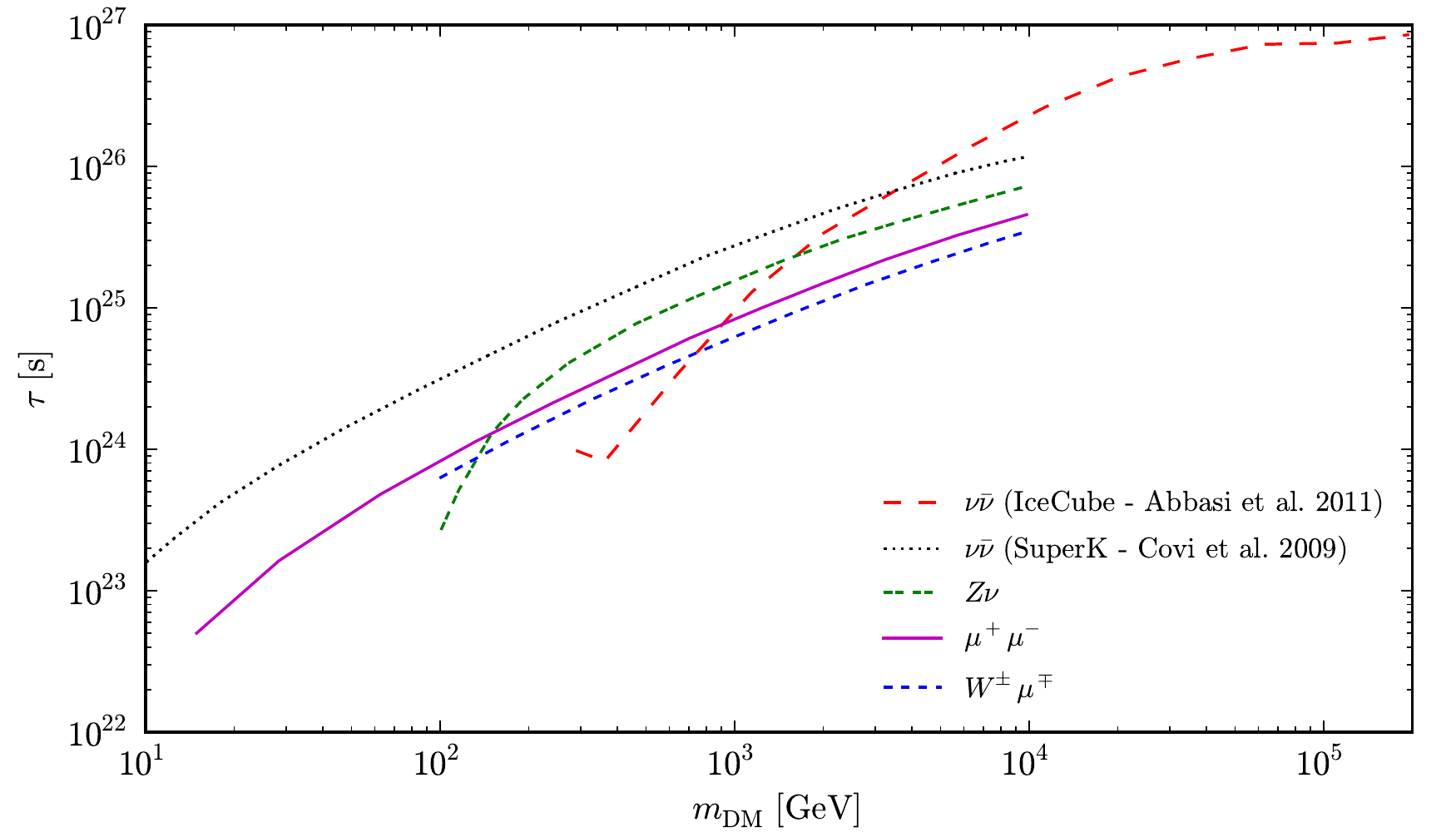}
  \caption{90\% C.L. exclusion region in the lifetime vs. mass plane for a
    decaying dark matter candidate from the non-observation of an excess in the
    Super-K data (from Ref.~\citen{Covi:2009xn}) and IceCube (from
  Ref.~\citen{Abbasi:2011eq}).}
  \label{SK-Icecube-limit}
\end{figure}

Several groups have searched for signals of dark matter decays in neutrino
telescopes
\cite{Covi:2008jy,Covi:2009xn,Hisano:2008ah,PalomaresRuiz:2007ry,Lee:2011nt,Abbasi:2011eq}.
No signal has been observed, thus implying lower limits on the dark
matter lifetime from the Super-Kamiokande data or the IceCube data, which are
shown in Fig.~\ref{SK-Icecube-limit} for various decay channels. The search
for neutrinos from dark matter decay has been extended to heavier dark matter
masses in Refs.~\citen{Esmaili:2012us,Murase:2012xs}, finding lower limits for
the lifetime of ${\cal O}(10^{26}-10^{28})\s$ for masses between 10 TeV and
the Grand Unification scale. 
 
\section{Models}
\label{sec:models}

\subsection{Lagrangian Analysis}
Before presenting some concrete models containing unstable dark matter
candidates, we will first introduce generic Lagrangians leading to the decay
of the dark matter
particles~\cite{Garny:2010eg,Esmaili:2012us,Gustafsson:2013gca}. Let us first
discuss the scenario where the dark matter particle is a scalar. In this case
the effective Lagrangian that induces the decay $\phi_{\rm DM}\rightarrow f
\bar f$ contains a dimension 4 operator:
\begin{equation}
  -{\cal L}_{\rm eff}=y \phi_{\rm DM} \bar f f+{\rm h.c.}
\end{equation}
where $y$ is the coupling constant. The decay rate can be straightforwardly calculated, the result being:
\begin{equation}\label{eq:lambda}
  \Gamma(\phi_{\rm DM}\rightarrow f \bar f)=\frac{|y|^2 N_c}{8\pi}m_{\rm DM}~.
\end{equation}
where $N_c$ is a color factor. For $N_c=1$ the lifetime is 
\begin{equation}
  \tau_{\rm DM}\simeq 2\times 10^{26}\s \left(\frac{y}{10^{-26}}\right)^{-2}
  \left(\frac{m_{\rm DM}}{1 \TeV}\right)^{-1}\;.
\end{equation}
Therefore, present cosmic rays measurements require a very large suppression of
the effective coupling if the dark matter particle is a scalar which decays into
a fermion--antifermion pair. A similar conclusion applies to a vector dark
matter particle that decays into a fermion--antifermion pair.

If the dark matter particle is a spin-1/2 fermion, the decays are
necessarily induced by higher dimensional effective operators generated by
dark matter interactions with heavy scalars or heavy vectors. A general
discussion of the possible three-body decays can be found in
Ref.~\citen{Garny:2010eg}. For example, the dark matter decay $\psi\rightarrow
f\bar f \nu$ can be induced by the following Lagrangian involving the heavy
scalar $\Sigma$:
\begin{align}
  \mathcal{L}_\text{eff}^\Sigma = -\bar{\psi}_{\rm DM} \left[\lambda_{f
  \psi}^L P_L + \lambda_{f \psi}^R P_R\right] f \, \Sigma^\dagger
  -\bar{\nu}\lambda_{f N}^R P_R  f
  \, \Sigma^\dagger + \text{h.c.}\;,
  \label{eq:Lag-scalar}
\end{align}
where $P_L = (1 - \gamma^5)/2$ and $P_R = (1 + \gamma^5)/2$ are the left- and
right-handed chirality projectors, respectively. In this case, the dark matter
decay width is given by:
\begin{equation}\label{eq:GammathreebodySigma}
  \Gamma(\psi_{\rm DM}\rightarrow f \bar f \nu)=
  \frac{|\lambda_{\rm eff}|^4 N_c}{3072\pi^3}\frac{m_{\rm DM}^5}{M_\Sigma^4}\;,
\end{equation}
where $|\lambda_{\rm eff}|^4=(|\lambda_{f \psi}^R|^2+|\lambda_{f \psi}^L|^2)
|\lambda_{f \nu}^R|^2$ and $N_c$ is a color factor. In contrast, if the decay
is mediated by a charged vector, the effective Lagrangian reads\footnote{We
assume here that the decay is dominated by the charged-current interaction; in
more generality the decay could also be mediated by a neutral current
interaction.}
\begin{equation}
  \mathcal{L}_\text{eff}^V = -\bar{\psi}_{\rm DM} \gamma^\mu
  \left[\lambda_{f \psi}^L P_L + \lambda_{f \psi}^R P_R\right] f \,
  V_\mu^\dagger - \bar{\nu} \gamma^\mu 
  \lambda_{f N}^R P_R  f \, V_\mu^\dagger + \text{h.c.}\;.
  \label{eq:Lag-vector}
\end{equation}
In this case the decay width is:
\begin{equation}\label{eq:GammathreebodyV}
  \Gamma(\psi_{\rm DM}\rightarrow f\bar f \nu)=
  \frac{|\lambda_{\rm eff}|^4 N_c}{768\pi^3}\frac{m_{\rm DM}^5}{M_V^4}\;,
\end{equation}
where, again, $|\lambda_{\rm eff}|^4=(|\lambda_{f \psi}^R|^2+|\lambda_{f
\psi}^L|^2) |\lambda_{f \nu}^R|^2$ and $N_c$ is a color factor. Note that the
decay width for the decay mediated by a vector is a factor of four larger than
the one mediated by a scalar, under the assumption that $M_V=M_\Sigma$ in
Eqs.~(\ref{eq:GammathreebodySigma}) and (\ref{eq:GammathreebodyV}). In the
case that the decay is mediated by a heavy scalar, the lifetime reads, for
$N_c=1$, 
\begin{equation}
  \tau_{\rm DM}\simeq 6\times 10^{25}\s \left(\frac{|\lambda_{\rm eff}|}{1}\right)^{-4}
  \left(\frac{M_\Sigma}{10^{15}\GeV}\right)^4 
  \left(\frac{m_{\rm DM}}{1\TeV}\right)^{-5}\;,
  \label{eq:lifetime-3body}
\end{equation}
while in the case that the decay is mediated by a heavy vector the lifetime is
a factor of four smaller. As apparent from this formula, the longevity of the
dark matter particle requires either very large masses for the mediators
($M_\Sigma\gtrsim 10^{15}\GeV$ for $|\lambda_{\rm eff}|\sim 1$ and $m_{\rm
DM}\sim 1 \TeV$) and/or very small couplings in the interaction
($|\lambda_{\rm eff}|\lesssim  10^{-12}$ for $M_\Sigma\sim 2\TeV $and $m_{\rm
DM}\sim 1 \TeV$).

Particularly interesting are those Lagrangians leading to the two-body decay
of the dark matter particle into one or two photons, namely
$\phi\rightarrow\gamma\gamma$, $\gamma Z$ or $\gamma h$ for scalar particles
and  $\psi\rightarrow\gamma \nu$ for spin 1/2 particles. There are many
possibilities for the effective interactions that induce the decay (see
Ref.~\citen{Gustafsson:2013gca} for a comprehensive discussion). For example,
the decay of a spin 1/2 dark matter particle $\psi\rightarrow \nu\gamma$ 
can be induced by the following effective
Lagrangian~\cite{Giunti:2008ve,Esmaili:2012us,Gustafsson:2013gca}
\begin{equation}
  {\cal L}=\frac{1}{2}\bar\psi_{\rm DM}\, \sigma_{\alpha\beta}(\mu +\epsilon\gamma_5)\,
  \nu F^{\alpha\beta}+{\rm h.c.}
  \label{eq:L-fermion-2body}
\end{equation}
where $F^{\alpha\beta}$ is the electromagnetic field strength tensor, while
$\mu$ and $\epsilon$ are the magnetic and electric transition moments,
respectively. If the dark matter and the neutrino have the same CP parities,
the magnetic transition moment vanishes, while when they have opposite CP
parities, the electric transition moment vanishes. In either case, the decay
rate can be cast as
\begin{equation}\label{eq:Gamma-fermion-2body}
  \Gamma(\psi_{\rm DM}\rightarrow \nu \gamma)=\frac{|\mu_{\rm eff}|^2}{8\pi} m_{\rm DM}^3\;,
\end{equation}
where we have defined an effective neutrino magnetic moment, $|\mu_{\rm
eff}|\equiv\sqrt{|\mu|^2+|\epsilon|^2}$. The corresponding lifetime reads in
this case:
\begin{equation}
  \tau_{\rm DM}\simeq 2\times 10^{26}\s\left(\frac{|\mu_{\rm eff}|}{10^{-29}\GeV^{-1}}\right)^{-2}
  \left(\frac{m_{\rm DM}}{1\TeV}\right)^{-3}\;,
\end{equation}
which requires a very suppressed effective magnetic moment in order to obey
the strong limits on the dark matter lifetime from gamma-ray line searches.

In specific models, the effective Lagrangian Eq.~(\ref{eq:L-fermion-2body}) is
generated by quantum effects. Concretely, the Lagrangians,
Eqs.~(\ref{eq:Lag-scalar},\ref{eq:Lag-vector}) generate such a magnetic moment,
which can be parametrized as~\cite{Garny:2010eg}:
\begin{equation}\label{eq:Mu-fermion-2body}
  |\mu_{\rm eff}|=\frac{q\, m_{\rm DM}\, |\theta_{\rm eff}|^2}{64\pi^2 M_\Sigma^2}~,
\end{equation}
where $q$ is the electric charge of the fermion, $M_\Sigma$ is the mass of 
the heavy scalar particle and $\theta_{\rm eff}$ is a combination of the 
couplings of the dark matter particle and the neutrino to the heavy 
scalar in the loop. In this case, the lifetime reads:
\begin{equation}
  \tau_{\rm DM}\simeq 7\times 10^{28}\s
  \left(\frac{\theta_{\rm eff}}{1}\right)^{-4}
  \left(\frac{M_\Sigma}{10^{15}\GeV}\right)^{4}
  \left(\frac{m_{\rm DM}}{1\TeV}\right)^{-5}\;.
\end{equation}

The  decay $\psi\rightarrow \gamma\nu$ generated radiatively by the
Lagrangians, Eqs.~(\ref{eq:Lag-scalar},\ref{eq:Lag-vector}) has a width which is
about two-three orders of magnitude smaller than the width of the tree level,
three-body decay $\psi\rightarrow f\bar f\gamma$ (see
eq.\ref{eq:lifetime-3body}). However, the suppressed decay rate could be
compensated by the stronger limits on this channel from gamma-line searches.
In fact, as as shown in Ref.~\citen{Garny:2010eg}, in some scenarios the
limits from searches for gamma-ray lines can be competitive with the limits
from electron/positron measurements (and, interestingly, without suffering
from propagation uncertainties). Future instruments, such as the Cerenkov
Telescope Array\cite{Consortium:2010bc} will offer complementary limits on
this class of scenarios. One can follow a similar reasoning for scenarios
where the dark matter decays into a quark--antiquark pair and a neutrino. An
analysis comparing the limits on the parameter space from gamma-ray lines and
from the non-observation of a significant excess in the PAMELA measurements of
antiproton-to-proton fraction was presented in Ref.~\citen{Garny:2012vt}.

\subsection{Gravitinos in $R$-parity Breaking Vacua}

A very well-studied candidate of decaying dark matter is the gravitino in
$R$-parity breaking vacua (see Refs.~\citen{Takayama:2000uz,
Buchmuller:2007ui, Ibarra:2007wg, Ishiwata:2008cu,
Ishiwata:2009vx, Covi:2008jy, Delahaye:2013yqa}).  The gravitino, if it is
the lightest supersymmetric particle (LSP), constitutes a very promising
candidate for the dark matter of the Universe.  The gravitino relic abundance from 
thermal processes is calculable using the supergravity formalism, the result being dependent on
the reheating temperature of the Universe, the gravitino mass and the gluino
mass~\cite{Bolz:2000fu}. It can be shown that the correct relic abundance can
be achieved for gravitino and gluino masses in the 100 GeV -- a few TeV range,
as expected in gravity-mediated supersymmetry breaking scenarios, when the
reheating temperature is around $10^{10}$ GeV, which is compatible with the
lower bound on the reheating temperature required by thermal
leptogenesis~\cite{Fukugita:1986hr} which is $T_R\gtrsim 10^9$
GeV~\cite{Davidson:2002qv,Buchmuller:2004nz}. Therefore, scenarios with a
gravitino in the mass range $m_{3/2}=100$ GeV -- a few TeV can accommodate both
baryogenesis via leptogenesis and supersymmetric dark matter.

Despite being very attractive, this picture is not free from problems. In most
analyses of supersymmetric scenarios, $R$-parity conservation is implicitly
imposed. If this is the case, the next-to-LSP (NLSP) can only decay
gravitationally into gravitinos and Standard Model particles with a lifetime
such that the NLSP is present at the time of Big Bang nucleosynthesis,
possibly jeopardizing the successful predictions of the standard scenario.
Indeed, this is the case for most supersymmetric scenarios with gravitino dark
matter. Namely, if the NLSP is a neutralino, its late decays into hadrons
can dissociate the primordial elements~\cite{Kawasaki:2004qu}, and if the
NLSP is a stau, its presence during BBN catalyzes the production of $^6$Li,
resulting in an abundance which is in conflict with
observations~\cite{Pospelov:2006sc}. 
 
One simple  solution to the problems induced by the NLSP in cosmology is to 
assume that $R$-parity is not exactly conserved~\cite{Buchmuller:2007ui}.
Then, the NLSP can decay into two Standard model particles well before the
onset of Big Bang nucleosynthesis, avoiding altogether the BBN constraints.
When $R$-parity is no longer imposed, the gravitino LSP is no longer stable, but
instead decays into Standard Model particles, for instance via
$\psi_{3/2}\rightarrow \nu \gamma$. For the range of $R$-parity breaking
parameters necessary to preserve the successful predictions of the standard
BBN scenario and to preserve the baryon asymmetry generated by leptogenesis,
the gravitino lifetime is predicted to be in the range $\tau_{3/2}=10^{26}
-10^{40}$ s.\cite{Buchmuller:2007ui} Therefore, the scenario of gravitino dark
matter in $R$-parity breaking vacua, proposed originally to provide a
consistent thermal history of the Universe  with supersymmetric dark matter,
thermal leptogenesis and successful Big Bang nucleosynthesis, also leads to
potentially  observable signatures in the cosmic-ray
fluxes~\cite{Ibarra:2007wg, Lola:2007rw, Ishiwata:2008cu, Covi:2008jy,
Ji:2008cq, Ishiwata:2009vx, Buchmuller:2009xv, Chen:2009ew, Bomark:2009zm,
Choi:2010jt, Delahaye:2013yqa}.

\begin{figure}
  \centering
  \includegraphics[scale=0.8]{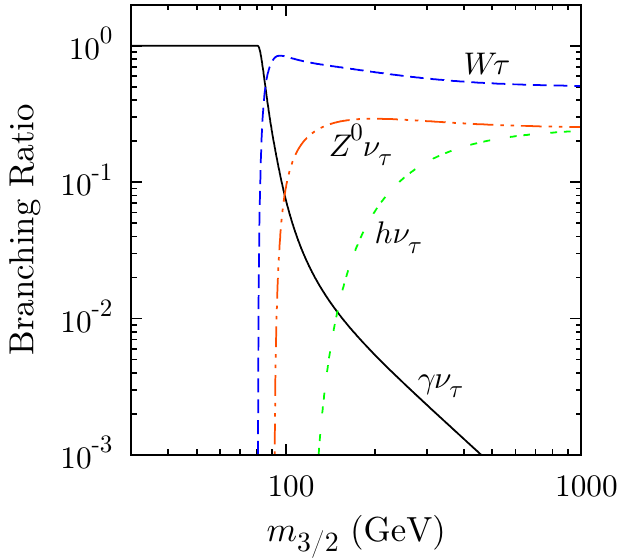}
  \caption{Branching ratios of the different gravitino decay channels as a
  function of the gravitino mass in a case of large Higgsino masses, gaugino
masses satisfying the unification relation and  giving  $M_1 = 1.5\, m_{3/2} $
at the electroweak scale and $\tan\beta = 10 $.}
  \label{fig:BR-grav}
\end{figure}

In models with bilinear $R$-parity breaking and a non-zero sneutrino vacuum
expectation value along the $\widetilde\nu_\tau $ direction, the main decay
channels for the gravitino are:
\begin{equation}
  \begin{split}
    \psi_{3/2} &\rightarrow\gamma \nu_{\tau}\,, \\
    \psi_{3/2} &\rightarrow W^{\pm} \tau^{\mp}\,, \\
    \psi_{3/2} &\rightarrow Z^0 \nu_{\tau}\,, \\
    \psi_{3/2} &\rightarrow h \nu_{\tau}\,.
  \end{split}
\end{equation}
The first decay is practically always allowed, while the remaining three only
above the production threshold of $W^\pm$, $Z^0$ or $h$. The decay widths for
these processes were calculated in Ref.~\citen{Covi:2008jy} and the three body
decays via a virtual weak gauge boson were studied in
Ref.~\citen{Choi:2010jt}. The branching fractions of the different gravitino
decay channels as a function of the gravitino mass are shown in
Fig.~\ref{fig:BR-grav}, for a case of large Higgsino masses, gaugino masses
satisfying the unification relation and  giving  $M_1 = 1.5\, m_{3/2} $ at the
electroweak scale and $\tan\beta = 10 $. As apparent from the plot, above the
$Z^0$ mass threshold the decay channels producing weak gauge bosons in the final 
state have large branching fractions, producing potentially observable signatures as an
excess in the cosmic antiproton-to-proton fraction. The non-observation of an
antiproton excess then allows to set a lower limit on the gravitino lifetime
which is ${\cal O}(10^{27}-10^{28})\s$ for
$m_{3/2}=100\GeV-1\TeV$~\cite{Delahaye:2013yqa}, with a strong dependence on
the antiproton propagation model. Complementary limits on this scenario follow
from the measurements of the gamma-ray flux by the Fermi LAT satellite,
concretely from observations of the isotropic gamma-ray background and of
galaxy clusters, which constrain the lifetime to be longer than $2\times
10^{26}\s$ for $m_{3/2}=100\GeV-1\TeV$. For lower masses, the limits from
searches of gamma-ray lines become fairly stringent and impose the lower limit
$\tau_{3/2}\gtrsim 5\times 10^{27}\s$ at
$m_{3/2}=100\GeV$.~\cite{Huang:2011xr}

\subsection{Hidden-Sector Gauginos}

Some extensions of the Minimal Supersymmetric Standard Model postulate a
hidden sector containing an extra unbroken Abelian gauge symmetry. In these
extensions, the hidden sector can communicate to the observable sector via the
kinetic mixing term of the hidden sector Abelian vector superfield with the
hypercharge vector superfield.\cite{Holdom:1985ag,Foot:1991kb,Dienes:1996zr}
If the extra gauge symmetry is unbroken, the
canonical normalization of the kinetic terms produces an unobservable shift of
the hypercharge gauge coupling and the generation of a milli-hypercharge for
the hidden sector chiral superfields. The existence of exotic particles with a
milli-hypercharge is severely constrained by experiments, although the limits
can be avoided when the masses of the exotic particles are large, which we
assume here. Furthermore, in the limit of unbroken supersymmetry, the
hypercharge vector superfield completely decouples from the observable sector
and is not subject to any experimental constraint. Nevertheless, the breaking
of supersymmetry dramatically changes the previous picture. Although the
hidden U(1) gauge boson remains decoupled from the observable sector, in the
presence of SUSY breaking effects a mixing between the hidden $U(1)$ gaugino,
$X$, and the MSSM neutralinos, $\chi_i^0$, is induced. In the presence of
R-parity conservation, a relic population of hidden gauginos can be generated
through the mixing with the MSSM neutralinos. The production of hidden
gauginos, as well as the limits on the parameters from primordial
nucleosynthesis and structure formation were discussed in
Ref.~\citen{Ibarra:2008kn}.

Depending on the masses of the hidden gaugino, $M_X$, and the lightest
neutralino, $M_{\chi_1^0}$, one of the two particles becomes unstable with a
lifetime roughly given by
\begin{equation}
  \tau_{X,\chi^0_1} \sim \mathcal{O}(10^{-2} - 10)\times 10^{26}\s  
  \cdot\left(
  \frac{M_{X,\chi^0_1}}{100\GeV} \right)^{-1} \left(
  \frac{\epsilon}{10^{-24}} \right)^{-2} \;,
  \label{eqn:lifetime}
\end{equation}
where $\epsilon$ is the kinetic mixing parameter. Such small values of the
kinetic mixing parameter naturally arise in specific models of
compactifications of heterotic and type II strings (see
e.g.~Ref.~\citen{Abel:2008ai}). The decay channels are
\begin{eqnarray}
  \chi_1^0 & \rightarrow & f\bar{f}X,\,X h^0,\, X Z^0 ~~~{\rm when~}
  M_{\chi_1^0}>M_X\;,
  \\
  X & \rightarrow& f\bar{f} \chi_i^0,\, \chi^0_ih^0, \,\chi^0_iZ^0, \,
  \chi^\pm_j W^\mp ~~~~{\rm when~} M_X>M_{\chi_1^0}\;,
\end{eqnarray}
where $f$ denotes any lepton or quark, which branching fractions which depend
on the concrete point of the MSSM parameter space. It is interesting to note
that the decay is dominated by the leptonic modes in certain parameter regions
of the MSSM where the sleptons are light. Thus, supersymmetric scenarios with a
hidden unbroken $U(1)$ gaugino can provide a leptophilic unstable dark matter
candidate. The cosmic-ray signatures of scenarios with hidden sector gauginos 
are discussed in Ref.~\citen{Ibarra:2009bm}.

\subsection{Hidden $SU(2)$ Vectors}

Scenarios with a hidden $SU(2)$ gauge group which is spontaneously broken
contain a natural dark matter candidate, the hidden vector $A^a_\mu$,
$a=1,2,3$, which is long-lived  due to an accidental custodial symmetry in the
renormalizable Lagrangian.\cite{Hambye:2008bq} Nevertheless,
non-renormalizable dimension-six operators, suppressed by the large mass scale
$\Lambda$, break the custodial symmetry and induce the decay of the dark
matter particle at cosmological times, see Eq.~\eqref{eq:target}. These are:
\begin{eqnarray}
\label{eqn:opA}
&{\rm (A)}&~~~\frac{1}{\Lambda^2}\  \mathcal{D}_{\mu}\phi^{\dagger}\phi\ \mathcal{D}_{\mu}H^{\dagger}H \,,\\
\label{eqn:opB}
&{\rm (B)}&~~~\frac{1}{\Lambda^2}\  \mathcal{D}_{\mu}\phi^{\dagger}\phi\  H^{\dagger}\mathcal{D}_{\mu}H \,,\\
\label{eqn:opC}
&{\rm (C)}&~~~\frac{1}{\Lambda^2}\  \mathcal{D}_{\mu}\phi^{\dagger}\mathcal{D}_{\nu}\phi\ F^{\mu\nu Y} \,,\\
\label{eqn:opD}
&{\rm (D)}&~~~\frac{1}{\Lambda^2}\ \phi^{\dagger}
F^a_{\mu\nu}\frac{\tau^a}{2}\phi F^{\mu\nu Y}\,,
\end{eqnarray}
where  $H$ the Standard Model Higgs doublet, $\phi$ is a complex $SU(2)_{\rm
HS}$ doublet scalar field which breaks the symmetry, $\mathcal{D}^\mu
=\partial^\mu \phi - i\frac{g_\phi}{2} \tau \cdot A^\mu$, with $\tau^a$,
$a=1,2,3$ the generators of the hidden $SU(2)$ gauge group, and $F_{\mu\nu}^Y$
and $F_{\mu\nu}^a$ are the field strength tensors of the hypercharge and the
hidden $SU(2)$ gauge group.

\begin{figure}
  \centering
  \includegraphics[width=0.49\linewidth,bb=0 0 350 223,clip]{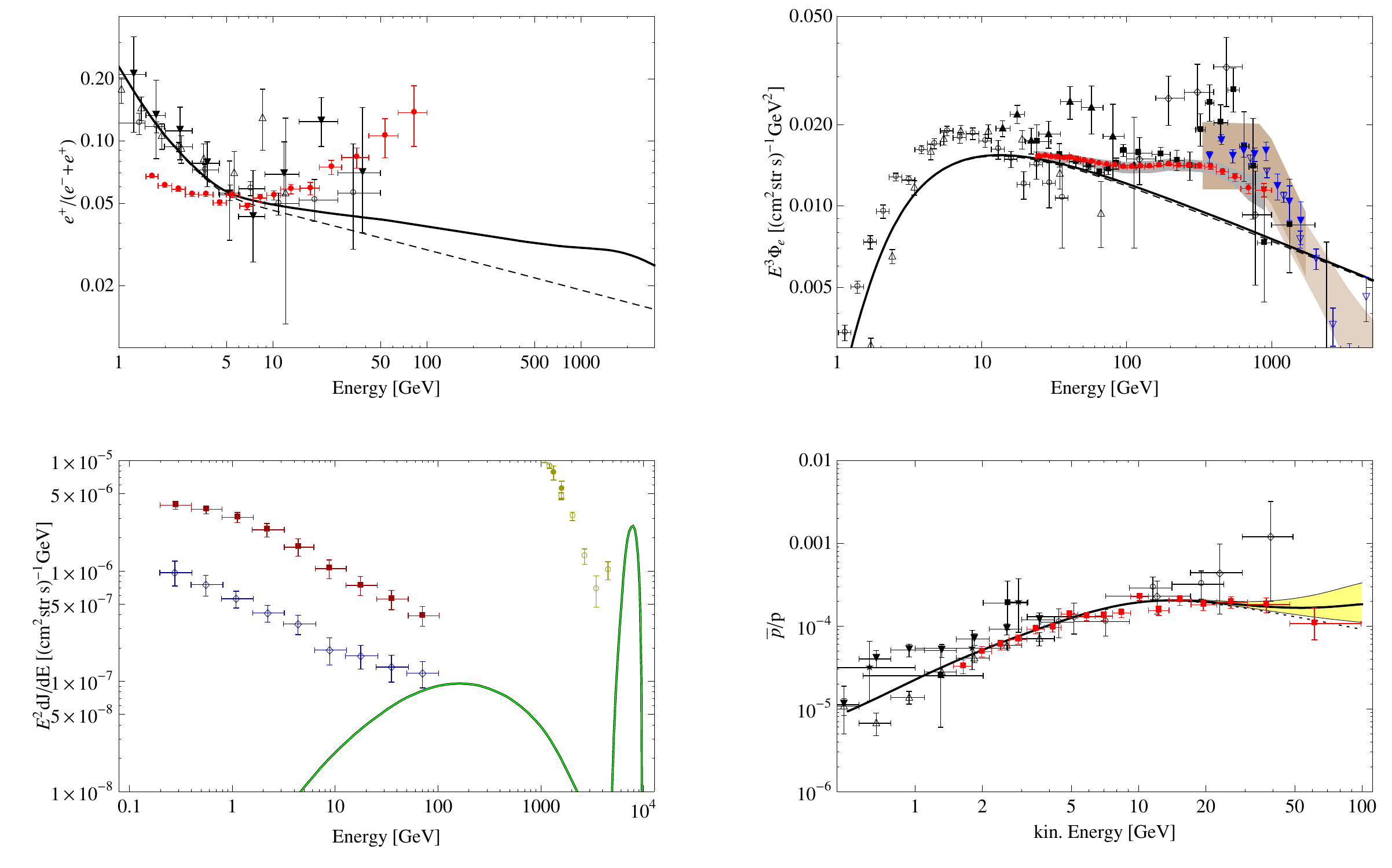}
  \includegraphics[width=0.49\linewidth,bb=0 0 350 223,clip]{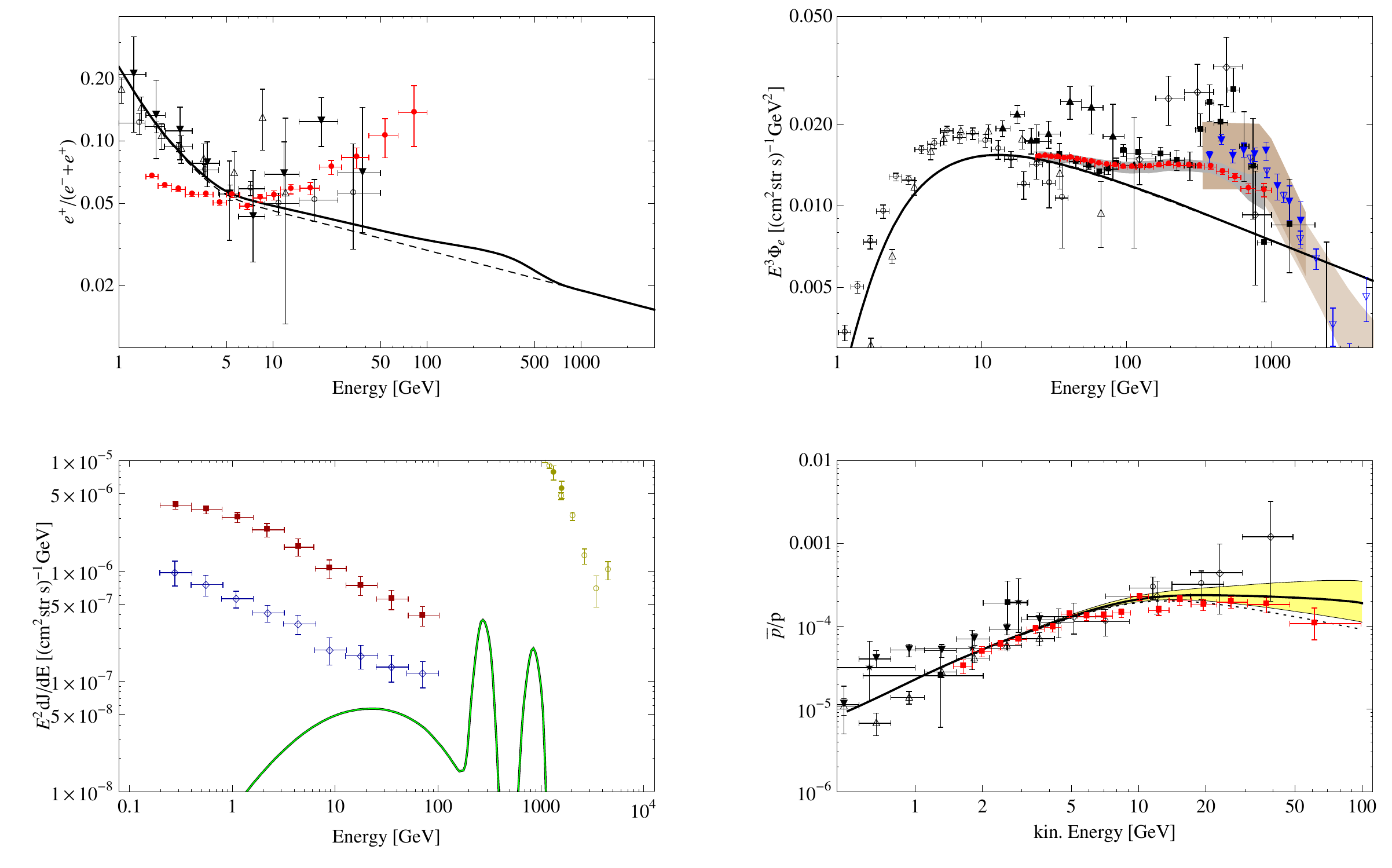}
  \caption{Isotropic gamma-ray signal from the decay of hidden $SU(2)$ vector
    dark matter particles, assuming that the decay is induced by the higher
    dimensional operator of the type (C), {\it cf.} Eq.~\eqref{eqn:opC}. The left
    plot corresponds to $\tau_{\rm DM}=6.0\times 10^{26}\s$ $(\Lambda=2.0\times
    10^{17}\GeV)$ and the right plot to $\tau_{\rm DM}=1.6\times 10^{27}\s$
  $(\Lambda=1.2\times 10^{16}\GeV)$. From Ref.~\citen{Arina:2009uq}.}
  \label{fig:SU(2)vector}
\end{figure}

The dark matter decay modes depend on which is the dominant operator breaking
the hidden sector custodial symmetry:
\begin{eqnarray}
&{\rm (A,B)}&~~~A\rightarrow \eta\eta, h\eta, hh, \gamma \eta, Z \eta, \gamma h, Z h \\
&{\rm (C)}&~~~A\rightarrow \gamma\eta,Z\eta,\gamma h, Z h \\
&{\rm (D)}&~~~A\rightarrow W^+ W^-,\, Z\eta, Zh, f \bar f, \gamma \eta, Z \eta, \gamma h, Z h
\end{eqnarray}
where $h$ and $\eta$ are mass eigenstates, linear combinations of the visible
and hidden sector Higgs bosons. In all cases, the decay of the hidden vector
dark matter produces two gamma-ray lines, from the decays $A\rightarrow \gamma
\eta$ and $A\rightarrow \gamma h$, with a rate that depends on which is the
dominant operator inducing the decay. For example, in the case C, in the limit
$M_\eta\ll M_A$, the inverse decay rate reads
\begin{equation}
  \Gamma(A\rightarrow \gamma\eta)^{-1} = 2.7\times10^{28}\s
  \left( \frac{\Lambda}{4\times10^{15}\GeV} \right)^4
  \left( \frac{300 \GeV}{M_A} \right)^5\;,
  \label{eqn:}
\end{equation}
which could be accessible to gamma-ray telescopes, depending on the scale of
custodial symmetry breaking and the dark matter mass; the rate for the decay
$A\rightarrow \gamma h$ is comparable.  In Fig.~\ref{fig:SU(2)vector} it is
shown the expected isotropic gamma-ray flux in two scenarios corresponding to
a decay induced by the operator of type (C). In the right plot one can clearly
see the presence of two intense gamma-ray lines. It is interesting that
present observations can, in this scenario, constrain the scale of the custodial
symmetry to be larger than the Grand Unification Scale.

\subsection{Right-Handed Sneutrinos in Scenarios with Dirac Neutrino Masses}

A simple extension of the Minimal Supersymmetric Standard Model consists of
introducing three right-handed neutrino superfields, thus allowing for
neutrino oscillations. As shown in
Ref.~\citen{McDonald:2006if,Asaka:2006fs}, the right-handed sneutrinos
constitute good cold dark matter candidates if $R$-parity is conserved and they
are lighter than the $R$-odd particles of the MSSM. While the $R$-parity
conservation ensures the absolute stability of the lightest supersymmetric
particle, namely the lightest right-handed sneutrino, the two heavier
right-handed sneutrinos could decay into the lightest and a lepton-antilepton
pair via neutralino/chargino exchange~\cite{Pospelov:2008rn}, $\tilde
\nu_{R1}\rightarrow \tilde \nu_{R2} \ell^+\ell^-$,  $\tilde
\nu_{R1}\rightarrow \tilde \nu_{R2} \nu\bar\nu$, provided the mass splitting
between the sneutrino eigenstates is large enough to kinematically allow the
decay. The decay rate for this process can then be approximated by
\begin{equation}
  \Gamma\simeq 10^{-6} y_1^2 y_2^2 m_1\;,
  \label{eq:sneutrino-DM}
\end{equation}
where $m_1$ is the mass of the decaying right-handed sneutrino and $y_1$
($y_2$) is the Yukawa couplings of the sneutrino $\tilde \nu_{R1}$ ($\tilde
\nu_{R2}$) with the neutralino/chargino and the Standard Model leptons (here
it was assumed for simplicity only one generation of leptons; the
generalization to three generations is straightforward). In scenarios with
Dirac neutrinos the size of the Yukawa couplings is expected to be
\begin{equation}
  y_\nu\simeq 3.0\times 10^{-13}\left(\frac{m_\nu^2}{2.8\times 10^{-3}\,{\rm
  eV}^2}\right)^{1/2}\;.
\end{equation}
Using this Yukawa coupling as reference, the lifetime of the
right-handed sneutrino next-to-LSP can then be cast as:
\begin{equation}
  \tau\simeq 8\times 10^{28} \s \left(\frac{y}{3.0\times
  10^{-13}}\right)^{-4}\left(\frac{m_ {\tilde \nu_{R1}}}{1\TeV}\right)^{-1}\;,
\end{equation}
which is remarkably close to the values that can be probed with 
cosmic ray observations. 

\subsection{Hidden Gauge Bosons}

This scenario considers an extension of the Standard Model 
by a hidden Abelian gauge symmetry
$U(1)_H$ and by a gauged $U(1)_{B-L}$ symmetry. It is further assumed that the
hidden sector matter does not interact directly with either the Standard
Model particles or with the $U(1)_{B-L}$; the only interaction arises through
the kinetic mixing term between the $U(1)_H$ and the $U(1)_{B-L}$ (this can be
arranged in an extra dimensional set-up, see
Refs.~\citen{Chen:2008md,Chen:2008qs}). It is also assumed that both the
$U(1)_H$ and the $U(1)_{B-L}$ are spontaneously broken, such that the
corresponding vector bosons acquire masses $m={\cal O}(100\GeV)$ and $M={\cal
O}(10^{15}\GeV)$ respectively. Then, the relevant Lagrangian reads:
\begin{eqnarray}
  {\cal L}&=& -\frac{1}{4}F^{(H)}_{\mu\nu}F^{(H) \mu\nu}-\frac{1}{4}F^{(B)}_{\mu\nu} F^{(B)\mu\nu}
  +\frac{\lambda}{2}F^{(H)}_{\mu\nu} F^{(B)\mu\nu}\nonumber\\
  &&+\frac{1}{2}m^{2} A_{H \mu}A_{H}^{\mu} +\frac{1}{2}M^{2} A_{B \mu}A_{B}^{\mu},
  \label{eq:kinetic}
\end{eqnarray}
where $A_H^\mu$ and $A_B^\mu$ are the vector bosons corresponding to the
$U(1)_H$ and $U(1)_{B-L}$ symmetries, while $F^{(H) \mu\nu}$ and $
F^{(B)\mu\nu}$ are the corresponding field strength tensors. In the absence of
a kinetic mixing term, the hidden gauge boson $A^H$ would be completely
stable. However, the kinetic mixing with the $U(1)_{B-L}$, $\lambda$, induces
the decay into the $B-L$ charged Standard Model fermions. The decay rate into
a fermion-antifermion pair with $B-L$ charge $q_i$ and color factor $N_i$ can
then be calculated to be
\begin{equation}
  \Gamma(A_H\rightarrow \psi_i \bar\psi_i)\simeq \lambda^2 \frac{N_i q_i^2}{12\pi}\frac{m^5}{M^4}\;.
\end{equation}
Hence the dark matter lifetime approximately reads:
\begin{equation}
  \tau \;\simeq\; \frac{2.5\times 10^{27} {\rm \,sec}}{\lambda^2 \sum_i N_i q_i^2}  \left(\frac{m}{100 \GeV}\right)^{-5}
  \left(\frac{M}{10^{15}\GeV}\right)^{4}\;.
\end{equation}
Note that the coefficient $N_i q_i^2$ is 1/3 for quarks and 1 leptons, hence
suppressing the antiproton flux. 

\subsection{Neutralino Decay in R-parity Breaking Vacua}

$R$-parity was introduced in the Minimal Supersymmetric Standard Model to
prevent too rapid proton decay. Introducing this discrete symmetry has the
appealing feature that the lightest neutralino, if it is also the lightest
supersymmetric particle, becomes absolutely stable and thus constitutes a
perfect candidate for  WIMP dark matter. However, there is no fundamental reason to assume
that $R$-parity should be completely exact and tiny $R$-parity breaking
parameters could leave observable signatures in cosmic rays, even for
parameters yielding a proton lifetime in agreement with
observations.\cite{Baltz:1997ar,Yin:2008bs,Shirai:2009fq,Sierra:2009zq}.  

In the framework of the Minimal Supersymmetric Standard Model, the most
general superpotential compatible with the gauge symmetry reads:
\begin{equation}
  W = W_{MSSM}  + \lambda_{ijk} L_i L_j \bar E_k + \lambda^\prime_{ijk}L_i Q_j \bar D_k + 
  \lambda^{\prime\prime}_{ijk}\bar U_i \bar D_j \bar D_k + \mu^\prime_i L_i H_u \;,
\end{equation}
where $i, j, k=1,2,3$ are generation indices, $\lambda$, $\lambda^\prime$,
$\lambda^{\prime\prime}$ are dimensionless $R$-parity breaking parameters and
$\mu^\prime$ has dimensions of mass. The lifetime of the lightest neutralino
must then be proportional to these $R$-parity breaking parameters. Namely, the
coupling $\lambda^\prime_{ijk}$ leads to the decays $\chi^0\rightarrow \nu_i
d_j\bar d_k$ and $\chi^0\rightarrow e^-_i u_j \bar d_k$, while
$\lambda^{\prime\prime}_{ijk}$ to $\chi^0\rightarrow u_i d_j d_k$ ($j\neq k$).
The corresponding lifetime can be estimated to be, in the limit of heavy and
degenerate sfermions and assuming only a non-vanishing coupling $\lambda^\prime$,
\begin{equation}
  \tau  \sim 10^{26} \s  \times \left( \frac{\lambda^\prime}{10^{ - 25}}
  \right)^{-2}  \left( \frac{m_\chi}{1 \TeV} \right)^{-1}\left(
  \frac{m_{\tilde f}}{m_\chi} \right)^4
\end{equation}
for a gaugino-like neutralino and  
\begin{equation}
  \tau  \sim 10^{26} \s \times \left(\frac{\tan \beta}{10} \right)^{-2}
  \left(\frac{\lambda^\prime }{10^{ - 23}}\right)^{-2}  \left(\frac{m_\chi
  }{1\TeV} \right)^{-1} \left(\frac{m_{\tilde f} }{m_\chi } \right)^4
\end{equation} 
for a Higgsino-like neutralino. In these expressions $m_{\tilde f}$ is the
common mass of the sfermions and $\tan\beta\equiv\langle H_u^0\rangle/\langle
H_d^0\rangle$ is the ratio between the expectation values of the two Higgs
doublets of the MSSM. The tiny $R$-parity violating coupling could result
from higher dimensional operators suppressed by a large powers of the scale
of new physics, as in certain scenarios involving flavor symmetries
(see Ref.~\citen{Sierra:2009zq}).

\subsection{Bound States of Strongly Interacting Particles}

The dark matter particle could be a composite state consisting of ``dark
quarks'' from a hidden sector or from the messenger sector of
supersymmetric models\cite{Hamaguchi:2008rv}, or techniquarks in technicolor
models \cite{Nardi:2008ix}. Quite generically, the Lagrangian contains higher-
dimensional operators, suppressed by a high energy scale, which induce the
dark matter decay. Therefore, the typical lifetime is given by
Eq.~\eqref{eq:target} which might in principle probed in cosmic ray
observations.

\subsection{Decaying Dark Matter from Dark Instantons}

The proton is the lightest baryon and therefore it is absolutely stable if the
vacuum is invariant under a global $U(1)$ baryonic symmetry. This symmetry is
accidental in the renormalizable part of the Lagrangian and could be broken or
not by higher dimensional operators. Nevertheless, the global $U(1)$-baryonic
is guaranteed to be broken by instanton $B+L$ violating effects which make the
proton unstable but very long-lived, with a lifetime $\approx 10^{140}$ 
years.\cite{'tHooft:1976up}.
This rationale was followed in Ref.~\citen{Carone:2010ha} to construct a
model where the dark matter stability is due to a global symmetry that is
broken only by instanton-induced operators generated by a non-Abelian dark
gauge group, $SU(2)_D\times U(1)_D$. The dark matter particle $\psi$ is in
this model predicted to decay $\psi\rightarrow \ell^+ \ell^- \nu$ with a rate
\begin{equation}
  \Gamma\approx
  \frac{1}{g_D^{16}}\exp(-16\pi^2/g_D^2)\left(\frac{m_\psi}{v_D}\right)^{47/3}m_\psi\;,
\end{equation}
where $g_D$ is the coupling constant and $v_D$ the expectation value of the
field that breaks the dark gauge group. Choosing $m_\psi=3.5$ TeV, $v_D=4 $
TeV and $g_D=1.15$ one finds a dark matter lifetime of $10^{26}\s$.


\section{Discussion and Conclusions}
\label{sec:conclusions}

While observations demonstrate that the dark matter particle is very long
lived, there is no astrophysical or cosmological observation requiring its
absolute stability. In fact, there are some well-motivated particle physics
scenarios which predict a long-lived dark matter particle, although not
absolutely stable. If the dark matter particle is indeed unstable, the decay
products could be observed as an excess in the cosmic-ray fluxes of antimatter
particles, gamma-rays or neutrinos over the expected backgrounds. In the last
few years a myriad of new experiments have provided data of exquisite quality
on the cosmic antimatter, gamma-ray and neutrino fluxes, which allow us to set
independent limits on the dark matter decay width into a given final state. In
this work we have focused on dark matter particles with masses in the
GeV--TeV range. We have reviewed the strategies to constrain the dark
matter decay width as well as the limits which follow from the most recent
observations.

The non-observation of an excess in the antiproton-to-proton fraction measured
by the PAMELA collaboration with respect to the expected astrophysical
backgrounds allows to set very stringent limits on the rate of decay processes
containing antiprotons in the final state. Concretely, the limits on the
lifetime for the decay ${\rm DM}\rightarrow W^+ W^-$ read
$\tau_{\rm DM}>20(3)\times 10^{27}\s$ and for the decay ${\rm
DM}\rightarrow b\bar b$,  $\tau_{\rm DM}>15(7)\times 10^{27}\s$, in both cases for $m_{\rm DM}=200(2000)\GeV$ 
and for the MED propagation model in Table \ref{tab:param-propagation}.
The positron fraction, however, shows an intriguing raise with the energy
which is at the moment not understood. Dark matter decays provide a
primary electron and positron flux which could account for the excess in
various decay channels, if the dark matter mass is $\sim$ few TeV and the
lifetime is $\sim 10^{26}\s$. Most of the models are however already in
tension with the non-observation of the associated inverse Compton emission in
Galaxy clusters, the isotropic gamma-ray background or the Galactic halo.  
No
positive measurement currently exists on the cosmic antideuteron flux, thus
allowing to set upper limits on the dark matter lifetime from antideuteron
searches. Decay modes producing antideuterons necessarily produce antiprotons,
hence there exists a strong correlation between the antiproton and the
antideuteron fluxes from dark matter decay. In fact, the limits on the dark
matter lifetime from the antiproton measurements are significantly stronger
than the limits from antideuteron measurements, and somewhat stronger than the
projected sensitivity of AMS-02 and GAPS. Therefore, the observation of
antideuterons from dark matter decay in these experiments will be challenging.

Indirect signals from dark matter decay exhibit less directional dependence
and amplification from regions of high dark matter density than those from
traditionally considered WIMPs, since the production rate is linear in the
dark matter density (as opposed to quadratic in the case of self-annihilating
WIMPs). As a consequence, the strongest gamma-ray constraints on the dark
matter lifetime come from observations of the extragalactic gamma-ray
background, whereas observations of the Galactic center are only weakly
constraining. Other important targets which yield similar sensitivity are nearby
galaxy clusters. Since the extragalactic gamma-ray background is isotropic, a
convincing identification of a dark matter contribution would have to mostly
rely on spectral information. Fortunately, gamma-ray lines, pronounced cutoffs
from final state radiation and box-shaped spectra -- that are predicted in
certain dark matter scenarios -- could facilitate such an identification.
No unambiguous dark matter signal has been found in gamma-ray observations,
yet.  If dark matter dominantly decays into $\bar b b$ or $\mu^+\mu^-$ final
states, and for dark matter masses between 10 GeV -- 10 TeV, Fermi LAT
observations of the extragalactic gamma-ray background and the Galactic halo
yield lower limits on the dark matter lifetime that range between $2\times
10^{25}\s$ and $2\times 10^{27}\s$, depending on the dark matter mass. In the
case of decay into $\gamma\nu$, much stronger constraints of the order
$3\times10^{29}\s$ and $10^{28}\s$ can be respectively derived from Fermi
(below 300 GeV) and HESS (above 500 GeV) dedicated line searches.

Neutrino searches are challenging because of the extremely small neutrino
interaction rate and the presence of large backgrounds, mainly from
atmospheric neutrinos. At present, no significant excess has been observed in
the diffuse neutrino fluxes, neither at Super-Kamiokande nor at IceCube,
allowing to set the limits on the lifetime $\tau_{\rm DM} \gtrsim 10^{25} \s$
for ${\rm DM}\rightarrow \nu\nu$,  $\tau_{\rm DM} \gtrsim 2\times 10^{24}\s$ for
${\rm DM}\rightarrow \mu^+\mu^-$,  $\tau_{\rm DM} \gtrsim 5\times 10^{23}\s$ for
${\rm DM}\rightarrow W^+W^-$ and  $\tau_{\rm DM} \gtrsim  5\times 10^{22}\s$ for
${\rm DM}\rightarrow b \bar b$, all of them at $m_{\rm DM}=2\TeV$.

These limits on the dark matter decay width set already stringent constraints
on the parameters of models containing decaying dark matter candidates. In
fact, some concrete scenarios proposed in the literature are ruled out by
present observations. The main implications of the
indirect searches for decaying dark matter for particle physics models are: 
{\it i)} for a scalar dark
matter particle decaying into a fermion-antifermion pair, the dimensionless
coupling inducing the decay must be $\lesssim {\cal O}(10^{-26})$ for $m_{\rm
DM}=1 \TeV$, {\it ii)} for a spin 1/2 particle decaying into a
fermion-antifermion pair and a neutrino mediated through the exchange of a
heavy scalar, the couplings involved in the decay must be tiny and/or the
mediator must be very heavy, concretely $M_\Sigma\gtrsim {\cal
O}(10^{15}\GeV)$ for $|\lambda_{\rm eff}|\sim 1$ or $|\lambda_{\rm
eff}|\lesssim  10^{-12}$ for $M_\Sigma\sim 2\TeV $, in both cases for $m_{\rm
DM}\sim 1 \TeV$; similar conclusions apply when the decay is mediated by a
vector, {\it iii)} for a spin 1/2 particle decaying into a photon and a
neutrino, the effective magnetic moment must be $\lesssim {\cal
O}(10^{-29}\GeV^{-1})$ for $m_{\rm DM}=1 \TeV$. Present indirect dark matter
searches are then sensitive to physics at very high energies, close to the
Grand Unification scale, or to very small couplings, possibly generated by
high dimensional operators or non-perturbative effects. To conclude, we note that 
the interesting role of cosmic-ray measurements in probing physics at very high 
energies is remarkable.

\section*{Acknowledgments}
The authors are grateful to Michael Grefe for useful comments.
A.I. and C.W.~thank the \emph{Kavli Institute for Theoretical Physics} in
Santa Barbara, California, for their kind hospitality. 
This work was partially supported by the DFG cluster of 
excellence ``Origin and Structure of the Universe.''

\appendix
\section{Analytical Solutions of the Transport Equation}
\label{apx:propagation}

\begin{table}[t]
  \tbl{Positron Green's function coefficients from Ref.~\citen{Ibarra:2008qg}.}
  {
    \begin{tabular}{ccc}
      \toprule
      Model & a & b \\
      \colrule
      M2 & $-0.9716$ & $-10.012$ \\
      MED & $-1.0203$ & $-1.4493$ \\
      M1 & $-0.9809$ & $-1.1456$\\
      \botrule
    \end{tabular}
  }
  \label{tab:positron-parameters}
\end{table}

\begin{table}[t]
  \tbl{Antiproton Green's function coefficients for the numerical
    approximation, Eq.~(\ref{eqn:antiproton_greensfunction_approx}).
  From Ref.~\citen{Ibarra:2008qg}.}
  {
    \begin{tabular}{cccc}
      \toprule
      Model & x & y & z \\
      \colrule
      MIN & $-0.0537$ & $0.7052$ & $-0.1840$ \\
      MED & $1.8002$ & $0.4099$ & $-0.1343$ \\
      MAX & $3.3602$ & $-0.1438$ & $-0.0403$\\
      \botrule
    \end{tabular}
  }
  \label{tab:antiproton-greensfunction}
\end{table}

\begin{table}[t]
  \tbl{Antideuteron Green's function coefficients, from Ref.~\citen{Ibarra:2009tn}.}
  {
    \begin{tabular}{cccc}
      \toprule
      Model & x & y & z \\
      \colrule
      MIN & $-0.3889$ & $0.7532$ & $-0.1788$ \\
      MED & $1.6023$ & $0.4382$ & $-0.1270$ \\
      MAX & $3.1992$ & $-0.1098$ & $-0.0374$\\
      \botrule
    \end{tabular}
  }
  \label{tab:antideuteron-greensfunction}
\end{table}

In this appendix we present semi-analytical solutions to the cosmic-ray
transport equation, Eq.~(\ref{eqn:transport}). Such solutions to the transport
equation can be found under certain simplifying assumptions by exploiting the
cylindrical geometry of the model. The first of these is that the diffusion
coefficient has no spatial dependence, $K(\vec{r}, T) = K(T)$ and can be
parametrized as
\begin{equation}
  K(T)=K_0 \,\beta\, {\cal R}^\delta\,.
\end{equation}

For positrons and electrons we can neglect diffusive reacceleration,
annihilation in the Galactic disk and convection. We further assume the
simplified form of the electron energy loss rate, $b(E)= \frac{1}{\tau_E}
\left(\frac{E}{E_0}\right)^2$.  As a result, we can describe electron/positron
propagation in terms of only three parameters which have to be determined from
observation, namely $\delta$, $K_0$ and $L$. We list three sets of parameters
in Table~\ref{tab:param-propagation}.  If we make the aforementioned
simplifications of spatially constant diffusion and energy loss rates, the
transport equation can be solved semi-analytically as a series in Bessel and
sin functions. The solution for the Green's function
is~\cite{Hisano:2005ec,Delahaye:2007fr}:
\begin{equation}
  G_{e^+}(T,T') = \sum_{n,m = 1}^\infty B_{nm}(T,T') J_0\left(\zeta_n
  \frac{r_\odot}{R}\right) \sin\left(\frac{m \pi}{2}\right)\;,
\end{equation}
where the coefficients are given by
\begin{align}
  B_{nm}(T,T') = & {} \frac{\tau_E T_0}{T^2} C_{nm} \times \nonumber \\
                 & \times \exp\left\{\left(\frac{\zeta_n^2}{R^2} + \frac{m^2
                 \pi^2}{4 L^2}\right) \frac{K_0 \tau_E}{\delta - 1}
               \left[\left(\frac{T}{T_0}\right)^{\delta - 1} -
             \left(\frac{T'}{T_0}\right)^{\delta - 1}\right]\right\} \,,
\end{align}
with
\begin{equation}
  C_{nm} = \frac{2}{J_1^2(\zeta_n) R^2 L} \int_0^R dr' \, r' \int_{-L}^L dz'
  \, \rho_\text{DM}(\vec{r}\,') J_0\left(\zeta_n \frac{r'}{R}\right)
  \sin\left[\frac{m \pi}{2 L} (L - z')\right] \,.
\end{equation}

For simplicity one can approximate the above Green's function numerically by a
rather simple parametrization instead of computing the full solution above.
The following simple function yields good numerical agreement with the
semi-analytical solution~\cite{Ibarra:2008qg}:
\begin{equation}
  G_{e^+}(T,T') \simeq \frac{10^{16}}{T^2} e^{a + b (T^{\delta - 1} -
  T'^{\delta - 1})} \theta(T' - T) ~ \text{cm}^{-3} ~ \text{s} \,.
\end{equation}
We list the coefficients $a$ and $b$ in Table~\ref{tab:positron-parameters}
for the NFW dark matter profile and the MIN, MED and MAX sets of transport
parameters.

For antiprotons and antideuterons we can neglect energy losses. We further
assume that the convective wind has axial direction away from the Galactic
disk. The analytic solution for the Green's function then reads
\begin{equation}
  G_{\bar{p}}(T,T') = \sum_{i = 1}^\infty \exp\left(-\frac{V_c L}{2
  K(T)}\right) \frac{y_i(T)}{A_i(T) \sinh(S_i(T) L / 2)}
  J_0\left(\zeta_i\frac{r_\odot}{R}\right) \delta(T - T')
\end{equation}
where
\begin{align}
  y_i(T) = {} & \frac{4}{J_1^2(\zeta_i) R^2} \int_0^R \, dr' \, r'
  J_0\left(\zeta_i \frac{r'}{R}\right) \int_0^L dz' \, \exp\left(\frac{V_c (L
  - z')}{2 K(T)}\right) \times \nonumber \\
              & \times \sinh\left(\frac{S_i (L - z')}{2}\right)
  \rho_\text{DM}(\vec{r}\,')
\end{align}
and
\begin{align}
  A_i(T) = & 2 h \Gamma_\text{ann} + V_c + k S_i(T) \coth\left(\frac{S_i(T)
  L}{2}\right) \,, \\
  S_i(T) = & \sqrt{\frac{V_c^2}{K(T)^2} + \frac{4 \zeta_i^2}{R^2}} \,.
\end{align}

The Green's function for antiproton propagation can be approximated
numerically by the following simple function~\cite{Ibarra:2008qg}:
\begin{equation}
  G_{\bar{p}}(T,T') \simeq 10^{14} e^{x + y \ln T + z \ln^2 T} \delta(T' - T)
  ~ \text{cm}^{-3}~\text{s} \,, \label{eqn:antiproton_greensfunction_approx}
\end{equation}
where the coefficients $x$, $y$ and $z$ are given in
Table~\ref{tab:antiproton-greensfunction} for the NFW dark matter profile and
the MIN, MED and MAX set of propagation parameters. The approximation is
better than 10\% compared to the full analytical solution. As mentioned
before, the choice of halo profile does not substantially affect the resulting
cosmic-ray fluxes in the case of decaying dark matter.  The case of
antideuteron propagation is almost identical to antiproton propagation and can
be approximated using the same parametrization,
Eq.~(\ref{eqn:antiproton_greensfunction_approx}).  The corresponding
parameters for the case of antideuterons are shown in
Table~\ref{tab:antideuteron-greensfunction}.

\bibliography{ddmr}
\bibliographystyle{ws-ijmpa}

\end{document}